\documentclass[journal]{IEEEtranTCOM}
\normalsize
\usepackage[pdftex]{graphicx}
  \usepackage{epstopdf}
  \DeclareGraphicsExtensions{.pdf,.jpeg,.png,.eps}
\usepackage{color}
\usepackage[cmex10]{amsmath}
\usepackage{amssymb}
\usepackage{algorithmic}
\usepackage{array}
\usepackage{mdwmath}
\usepackage{amsthm}
\usepackage{multirow}
\usepackage{mdwtab}
\usepackage{tabularx}
\usepackage{stfloats}
\usepackage{balance}
\usepackage{cite}
\makeatother
\allowdisplaybreaks
\newtheoremstyle{mystyle}
  {3pt}
  {3pt}
  {\itshape} 
  {\parindent}
  {\bfseries}
  {\upshape{:}}
  {.5em}
  {}

\usepackage{color,soul}

\theoremstyle{mystyle}

\theoremstyle{mystyle}  

\theoremstyle{mystyle}
\newtheorem{prop}{Proposition}
\newtheorem{lem}{Lemma}
\usepackage[british,UKenglish]{babel}
\usepackage{subcaption}
\usepackage[font={small}]{caption}
\graphicspath{{./figures/}}

\begin{document}

\title{Battery Recharging Time Models for Reconfigurable Intelligent Surface-Assisted Wireless Power Transfer Systems}
\author{Lina Mohjazi,~\IEEEmembership{Member,~IEEE,} Sami Muhaidat,~\IEEEmembership{Senior Member,~IEEE,} Qammer H. Abbasi,~\IEEEmembership{Senior Member,~IEEE,} Muhammad Ali Imran,~\IEEEmembership{Senior Member,~IEEE,} Octavia A. Dobre,~\IEEEmembership{Fellow,~IEEE,} \\ and Marco Di Renzo,~\IEEEmembership{Fellow,~IEEE}  
\thanks{L. Mohjazi is with the  School of Engineering, University of Glasgow, Glasgow, UK (e-mail: l.mohjazi@ieee.org).}
 \thanks{S. Muhaidat is with the Center on Cyber-Physical Systems, Khalifa University, Abu Dhabi, UAE, and also with the Department of Electrical Engineering and Computer Science, Khalifa University, Abu Dhabi, UAE (e-mail: muhaidat@ieee.org).} 
 \thanks{Q. H. Abbasi and M. A. Imran are with the School of Engineering, University of Glasgow, Glasgow, UK (e-mail: $\{\text{Qammer.Abbasi, Muhammad.Imran}\}\text{@glasgow.ac.uk}$).}
\thanks{O. A. Dobre is with the Department of Electrical and Computer Engineering, Memorial University, St. John’s, Canada (e-mail: odobre@mun.ca).}
\thanks{M. Di Renzo is with Universit$\acute{\text{e}}$ Paris-Saclay, CNRS, CentraleSup$\grave{\text{e}}$lec, Laboratoire des Signaux et Syst$\grave{\text{e}}$mes, Gif-sur-Yvette, France (e-mail: marco.direnzo@centralesupelec.fr).
The work of M. Di Renzo was supported by the European Commission through the H2020 ARIADNE Project under Grant 871464.}}

\maketitle
\markboth{}{}
\begin{abstract}
In this paper, we develop an analytical framework for the statistical analysis of the battery recharging time (BRT) in reconfigurable intelligent surfaces (RISs) aided wireless power transfer (WPT) systems. Specifically, we derive novel closed-form expressions for the probability density function (PDF), cumulative distribution function, and moments of the BRT of the radio frequency energy harvesting wireless nodes. Moreover, closed-form expressions of the the PDF of the BRT is obtained for two special cases: i) when the RIS is equipped with one reflecting element (RE), ii) when the RIS consists of a large number of REs. Capitalizing on the derived expressions, we offer a comprehensive treatment for the statistical characterization of the BRT and study the impact of the system and battery parameters on its performance. Our results reveal that the proposed statistical models are analytically tractable, accurate, and efficient in assessing the sustainability of RIS-assisted WPT networks and in providing key design insights for large-scale future wireless applications. For example, we demonstrate that a 4-fold reduction in the mean time of the BRT can be achieved by doubling the number of RIS elements. Monte Carlo simulation results corroborate the accuracy of the proposed theoretical framework.

\end{abstract}
\section{Introduction}
\IEEEPARstart{T}{he} roadmap to beyond the fifth generation (B5G) wireless networks is envisaged to introduce a new spectrum of fully automated and intelligent data-driven services, such as flying vehicles, haptics, telemedicine, augmented and virtual reality \cite{di2019smart,8796365,Liaskos1, 8910627, bariah2020prospective}. Several  unprecedented application environments, including machine-to-people and machine-to-machine communications, are expected to be the driving force of B5G systems. As a result, the number of connected Internet-of-Everything (IoE) devices (e.g. sensors, wearables, implantables, tablets) is anticipated to witness a phenomenal growth in the next few years, reaching up to tens of billions \cite{tariq2019speculative}. This poses a fundamental challenge on provisioning a ubiquitous seamless connectivity, while concurrently prolonging the lifetime of a massive number of energy-constrained low-power low-cost devices.
\par Wireless power transfer (WPT) has been highly recognized in both academia and industry as a promising technology to address the energy sustainability problem of wireless nodes, and has rapidly gained a growing interest in the research of B5G communication networks \cite{Varshney2008,Grover}. This is mainly due to its capability to deliver on-demand wireless energy to a large number of wireless devices in a controllable and low-cost manner, and thus, eliminating the need for battery replacement. In this framework, radio frequency (RF) signals, that are received from dedicated wireless power transmitters, are leveraged either to recharge the batteries of wireless nodes or to directly power the wireless transmissions of battery-less devices \cite{Bi,Mohjazi2}. 
\par However, it is demonstrated in recent studies \cite{8539989,8411158,Nasir2013} that the distance between the RF transmitter and the corresponding RF energy harvesting (RFEH) receiver creates a performance bottleneck for practical wirelessly-powered wireless networks. This stems from the fact that the efficiency of WPT is inversely proportional to the distance, and hence, conventional relay-aided wireless communications were proposed to realize WPT range expansion \cite{8539989,8411158,Nasir2013}. More recently, advanced technologies, such as massive multiple-input multiple-output (MIMO), are studied as potential candidates to achieve considerable WPT efficiency gains through the exploitation of beamforming with or without relaying techniques \cite{8353836,7374738}. Nonetheless, this comes at the price of severe energy consumption, higher computational complexity, and increased hardware cost, all of which are more pronounced at higher RF frequencies, such as millimeter-wave frequencies for future wireless systems \cite{8680660}. In addition to the distance factor, the efficiency of WPT, achieved through these conventional mechanisms, is degraded by the RF signal attenuation resulting from high penetration loss, multi-path fading, molecular absorption, and Doppler shift \cite{Liaskos1}. This effect is even more noticeable in ultra-dense network deployments, which feature highly dynamic radio environments. 
\par To address the afore-mentioned challenges, \textit{reconfigurable intelligent surfaces} (RISs) have recently emerged as a promising technology that can potentially offer fundamental performance improvements in wireless systems, in terms of spectrum and energy efficiencies, in a cost-effective manner \cite{di2019smart,8796365,Liaskos1,mohjazi2020outlook,yang2016programmable,huang2019holographic}. Practically, RISs can be realized by different approaches. This includes (i) surfaces equipped with large arrays of discrete, inexpensive, and tiny antenna elements, called unit cells, (ii) implementations based on conformal large surfaces or metamaterial-based planar with scattering elements spaced apart at distances much smaller than the wavelength \cite{di2020smart}. Unlike conventional approaches that lack full control over the propagation environment, RISs allow a transformative control of electromagnetic (EM) waves \cite{8910627,di2019smart,di2020smart}. For example, in RISs made of large arrays of inexpensive antennas, each element is individually configured and optimized to manipulate the impinging EM waves in arbitrary ways \cite{dunna2020scattermimo,246282}. This is achieved by jointly manipulating the reflected signal amplitude and/or phase shift at each of the RIS elements in real time according to the dynamic and implicit randomness of wireless channels. For instance, the signal component arriving from an RF source node, and reflected by the RIS elements, can be steered towards an intended destination node to enhance its received signal power. Traditional transmission techniques, such as phased arrays, multi-antenna transmitters, and relays, involve active components of complex hardware that exhibit high power consumption. On the other hands, RISs require a large number of scattering elements, each of them is supported by the lowest number of small-sized, low-power, and inexpensive components \cite{dunna2020scattermimo,246282}. Details of the key similarities and differences between RIS and relays are provided in \cite{ntontin2019reconfigurable}. 
\vspace{-0.1cm}
\subsection{Related Work}
The opportunities opened by RIS have spurred, in a short span of time, research in many areas related to wireless communication systems. This includes multi-user resource allocation, beamforming optimization, design of efficient enabling mechanisms, and performance analysis of RIS-assisted wireless networks. For example, in the area of resource allocation, the authors in \cite{huang2019reconfigurable} developed energy-efficient power allocation approaches  subject to individual link budget guarantees for multiple mobile users. The study in \cite{9014204} proposed an achievable rate optimization framework for orthogonal frequency division multiplexing (OFDM) that jointly identifies the transmit power at the base station (BS) and the reflection coefficients at the RIS. Furthermore, several research works designed enabling mechanisms, such as channel estimation schemes in an effort to achieve the passive beamforming gains of RIS \cite{9087848,zappone2020overhead}. Also, to reduce the overhead in channel training, the authors in \cite{9039554} proposed a practical transmission protocol that involves estimating the combined channel of a group of RIS elements. Furthermore, the authors in \cite{8981888} and \cite{zhang2018space} investigated the realization of index modulation and space-time modulated digital coding, respectively, in RIS-assisted transmissions to improve the spectral efficiency. From the performance analysis point of view, recent research studies provided a theoretical framework to characterize the performance of RIS-assisted wireless systems in terms of outage probability \cite{jung2019reliability}, reflection probability \cite{di2019reflection}, spectral efficiency \cite{9013789, zhou2020spectral}, and capacity \cite{zhang2019capacity,karasik2019beyond,perovic2019channel}. Additionally, the fundamental limits of the error probability performance of RIS-aided backscatter and non-orthogonal multiple access (NOMA) were, respectively, examined in \cite{zhao2020performance} and \cite{9027303}. Meanwhile, numerous research studies focused on proposing active and passive beamforming strategies to achieve secrecy \cite{8847342} and sum-rate \cite{zhang2020sum} enhancements and transmit power reduction \cite{8811733,han2019intelligent,zhou2019robust} for multi-antenna and/or multi-user RIS-assisted networks. 
\par While the preceding research works focus on leveraging the benefits of RIS to improve the performance of information transfer, RIS is shown to offer significant enhancements in the power transfer efficiency in wireless systems powered by near-field WPT \cite{8000613}. Apart from this, the deployment of RIS is demonstrated to enhance the far-field WPT and establish effective RFEH zones through compensating RF signals over long distances \cite{8941080,pan2020intelligent,wu2019joint}. Specifically, the work in \cite{8941080} and \cite{pan2020intelligent} considered the weighted sum-power and sum-rate optimization problems, respectively, for RIS-aided simultaneous wireless information and power transfer (SWIPT) systems, where an RIS is deployed in the vicinity of two separate groups of energy and information receivers. Moreover, joint active and passive beamforming design for RIS-aided SWIPT is proposed in \cite{wu2019joint} taking into consideration the signal-to-interference-plus-noise ratio constraints imposed by the information receivers (IRs) and RFEH constraints imposed by the energy receivers (ERs).
\vspace{-0.1cm}
\subsection{Motivation and Contribution}
The afore-mentioned studies in \cite{8941080,pan2020intelligent,wu2019joint} assume battery-free RFEH ERs, whose harvested energy is directly used for future transmissions. In this case, the amount of the received RF signals, and consequently, the amount of harvested energy, is considered to be sufficient and predictable over a certain period of time. However, in scenarios where RFEH nodes are equipped with batteries \cite{6952122}, the harvested energy is stored first in the battery before being used for future transmissions. Since the power of the received RF signal depends on the distribution of the probabilistic wireless fading channel between the transmitter and the receiver, the RFEH process and similarly, the time required to recharge the battery of an RFEH node, called the battery recharging time (BRT), become stochastic processes. The statistical characterization and modeling of BRT were developed in \cite{6952122,7374745,8269106} for conventional SWIPT systems operating over multi-path fading channels. Their results demonstrated that the RFEH process is significantly impacted by the system, fading, and battery parameters, including the capacity and discharging depth, and the BRT. Although the previously mentioned research work, carried out in \cite{8941080,pan2020intelligent,wu2019joint}, provide useful results to improve the RFEH process in RIS-assisted SWIPT systems, their approach focuses on developing transmission protocols and their insights are limited to battery-free ERs. 
\par Despite being a fundamental figure of merit in designing and quantifying the sustainability of RIS-assisted WPT networks in various operational setups in B5G systems, to the best of our knowledge, the statistical characterization of BRT is not yet studied in the open literature. Motivated by this, the prime focus of this work is to develop a novel theoretical framework to characterize the statistical properties of BRT for RIS-assisted WPT systems, consisting of ERs with limited battery capacity. In our work, we consider that RIS comprises passive REs spaced half of the wavelength apart, and that each element is individually configured for realizing optimal WPT. The main contributions of this paper are listed in more details as follows:
\begin{itemize}
    \item First, we derive novel accurate closed-form approximations for the  probability density function (PDF) and cumulative distribution function (CDF) of the instantaneous total received power of the energy harvesting node. The obtained expressions take into account the number of RIS REs and the distances of the source (S)$\to$RIS and RIS$\to$ER links, and assume that all wireless channels are subject to Rayleigh fading.
    \item Building upon the received power analytical expressions, we obtain closed-form expressions for the PDF and CDF of the BRT. The derived results are shown to accurately capture the impact of the battery parameters (i.e., discharge depth, battery capacity, and charging voltage), system parameters, and the number of RIS elements on the BRT statistical properties. To the best of our knowledge, these expressions are novel in literature.
    \item Next, analytical closed-form expressions for the PDF of the BRT are derived for two special cases, namely 1) when the RIS is equipped with only one RE, which serves as a benchmark to quantify the gains obtained by increasing the number of RIS elements and 2) when the number of REs grows large. For the later case, we exploit the central limit theorem (CLT) to demonstrate that the PDF of the BRT converges to an impulse response, revealing that the deployment of RIS is highly promising in practically realizing WPT in future large-scale IoE networks.
    \item To further investigate the properties of the BRT, we derive a computationally simple closed-form expression for its moments. We employ this result to obtain statistical tools to evaluate the mean value, variance, skewness and kurtosis of the BRT.
    \item The derived moments expression is also exploited to examine the amount of fading (AoF) of the BRT
as a function of RIS elements. Our study unveils that employing RISs of large size can effectively boost the efficiency of WPT over fading channels.
    \item Finally, we present Monte Carlo simulation and numerical results to validate the accuracy of the developed theoretical framework. 
\end{itemize}
\vspace{-0.2cm}
\subsection{Organization}
The rest of the paper is organized as follows: Section~\ref{model} provides the RIS-assisted WPT system model, as well as the statistical characterization of the corresponding power received by the energy harvesting node. In Section~\ref{BRTstat}, the analytical expressions of the statistics of the BRT for RIS-assisted WPT systems are derived. Simulation and numerical results are presented in Section~\ref{results}, while concluding remarks are given in Section~\ref{conc}. 

 
\section{System and Channel Model}\label{model}
\begin{figure}[ht]
\centering
\includegraphics[width=3.5in]{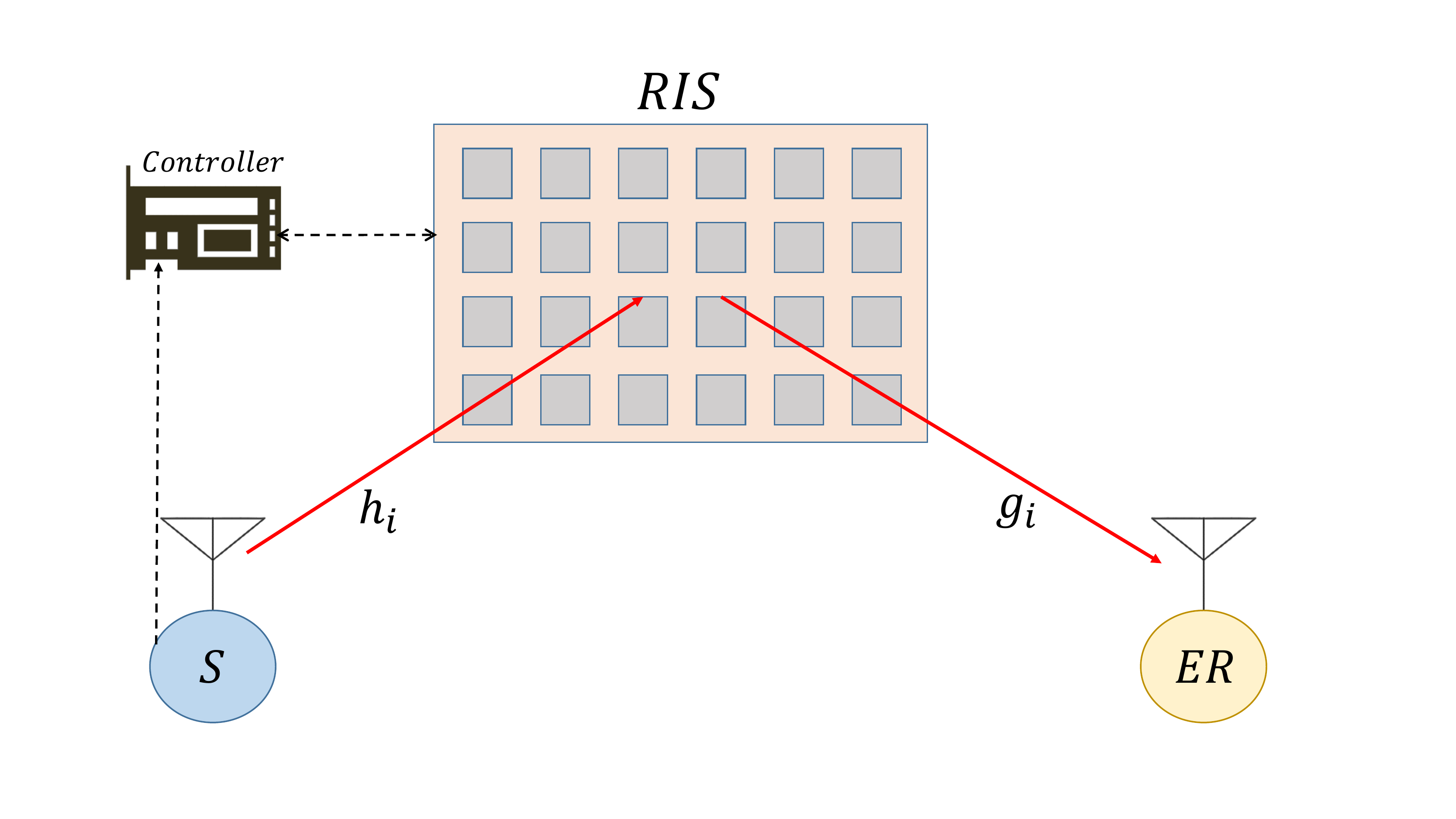}
\caption{RIS-assisted WPT system model.}
\label{systemmodel}
 \end{figure}

In this paper, we consider a single-antenna RF source node, $S$, and a single-antenna energy-constrained ER, as depicted in Fig.~\ref{systemmodel}. The ER could be a low-power sensor node equipped with a battery with a finite capacity. In order to extend the operational range of the ER, while ensuring that its harvested energy is sufficient for real-life operation, we propose WPT assisted by an RIS.
\par  The end-to-end (E2E) channel gain between $S$ and ER characterizes the power received at the ER, and accordingly, defines the behavior of the overall RFEH process, including the instantaneous BRT at ER. Therefore, to quantify the impact of RIS-assisted WPT on the required time to charge the battery of the ER node, in this section, we analytically present the distribution of the instantaneous received power at ER, which will be exploited next to develop the statistical characterization of the underlying instantaneous BRT.
\par We further assume that a direct link does not exist between $S$ and ER. This is motivated by the fact that this link is subject to  strong attenuation, due to deep fading, or shadowing effects, due to surrounding physical obstacles, or both and WPT can be achieved only via the RIS.  It is worth mentioning that such an assumption is widely adopted in research studies related to WPT communication systems \cite{8411158,pan2020intelligent}. 
\par In our setup, we consider an RIS that is made of $\textit{N}$ REs. Each of the elements can be reconfigured by a communication oriented software through a controller, as illustrated in Fig.~\ref{systemmodel}. The power transmitted from $S$, being either a BS or an RF source, which is reflected by the RIS towards the ER, is harvested and stored in a battery with a limited capacity before being used in future signal transmissions. 
\par  As shown in the block diagram of Fig.~\ref{systemmodel}, $h_i$ and $g_{i}$ denote the small scale complex channel fading coefficients of the $S$ $\to$RIS and RIS$\to$ER links, respectively, where $i\in \{1,2,...,N\}$ denotes the $i$-th element of RIS. It is assumed that the envelopes of the two wireless links are modeled as independent and identically distributed (i.i.d) Rayleigh fading channels with scale parameter, $\sigma$, being equal to 1, i.e., $|h_i|$, $|g_i|$$\sim$$\mathcal{C}\mathcal{N}(0,2\sigma^2)$ for $i$$\in$ $\{ 1, 2, . . . ,N\}$, where $\mathcal{C}\mathcal{N}(0,\kappa)$ stands for a zero-mean complex Gaussian distribution with $\kappa$ variance. The assumption of Rayleigh fading channels is representative of scenarios in which line-of-sight (LOS) propagation cannot be established due to random RIS deployments, e.g., if the RISs are deployed on spatial blockages. In this case, the system designer has no control over optimizing RISs locations \cite{qian2020beamforming}.

 Let $P_s$ denote the transmit wireless power of the source node. The instantaneous total power received at ER through the $i$-th element of RIS is expressed as
\begin{equation}\label{Pr-D}
 P_{r}= \frac{\left|\sum_{i=1}^N |h_i| |g_{i}| e^{j\theta_i}\right|^2}{d_{1}^\delta d_{2}^\delta} P_s,
\end{equation} where $d_{1}$ and $d_{2}$ represent the distance between $S$ and the center of the RIS and between the center of the RIS and ER, respectively, and $\delta$ is the path loss exponent. Furthermore, $\theta_i$ specifies the adjustable phase induced by the $i$-th RE of the RIS, respectively  \cite{8796365}. The total received power expression given in \eqref{Pr-D} is applicable in the far-field regime, as defined in \cite{tang2019wireless,di2020analytical}. Accordingly, $N$ can be large but it needs to be finite \cite[Sec.~IV-D]{di2020smart}. 
\par It is assumed that the channel phases of $h_i$ and $g_{i}$, denoted respectively as $\phi_{h_i}$ and $\phi_{g_{i}}$,  are perfectly known to the RIS and accordingly, it is able to provide optimal phase shifting, i.e., $\theta_i=-(\phi_{h_i}+\phi_{g_{i}})$. This idealized scenario sets a system operation performance benchmark for practical operations \cite{8796365}. Therefore, the results obtained represent a lower bound on the BRT required to sustain the operation of RIS-assisted WPT systems. In this paper, we assume that the amplitude of the reflection coefficient of each RE is equal to 1,\footnote{Recent studies proposed efficient designs achieving a reflection coefficient value as high as 1 \cite{8910627}.} for all $i\in N$. Consequently, the instantaneous received power is maximized and \eqref{Pr-D} can be formulated as
\begin{equation}\label{Pr-D2}
 P_{r}= \frac{P_s}{d_{1}^\delta d_{2}^\delta} B^2
\end{equation} where 
\begin{equation}\label{B}
B=\sum_{i=1}^N |h_i| |g_{i}|    
\end{equation} is the E2E channel gain.
\subsection{Statistical Characterization of the E2E Channel}
%
%
\par \noindent As previously mentioned, the BRT, $T_r$, is determined by the amount of power received and then harvested at ER. Therefore, it is necessary to have in hand the statistical characterization of the E2E channel fading coefficient, $B$, in order to derive the distributions of $P_{r}$ and $T_r$. Note that $B$ presents a sum of $N$ double Rayleigh random variables (RVs). An accurate approximation of its PDF is delivered in the following proposition.
\newcounter{tempequationcounter}
\begin{figure*}[ht]
\normalsize
\setcounter{equation}{15}
\begin{align}\label{mu_444}
\mu_{4}=\left\{ \begin{array}{l}
\left(64N+48N(N-1)+9N(N-1)\pi^{2}+6N(N-1)(N-2)\pi^{2}+\frac{N(N-1)(N-2)(N-3)\pi^{4}}{16}\right),N\geq4\\
\left(480+90\pi^{2}\right),N=3\\
\left(224+18\pi^{2}\right),N=2\\
64,N=1.
\end{array}\right.
\end{align}
\hrulefill
\end{figure*}
\setcounter{equation}{3}
\begin{prop} 
The PDF of the E2E channel coefficient of RIS-assisted WPT is accurately approximated in a closed-form as
\begin{equation}
f_{B}(x)\approx a_{1}G_{1,2}^{2,0}\left[\frac{x}{a_{2}}\left\vert \begin{array}{c}
-;a_{3}\\
a_{4},a_{5};-
\end{array}\right.\right], x\geq0  \label{ccdf1}
\end{equation}where $ G^{.,.}_{.,.}[.\vert .]$ denotes the Meijer G-function defined in \cite[Eq. (8.2.1.1)]{Prudnikov} and 
\begin{equation}
a_{1}=\frac{\Gamma(a_{3}+1)}{a_{2}\Gamma(a_{4}+1)\Gamma(a_{5}+1)},\label{a1}
\end{equation}
\begin{equation}
a_{3}=\frac{4\varphi_{4}-9\varphi_{3}+6\varphi_{2}-\mu_{1}}{-\varphi_{4}+3\varphi_{3}-3\varphi_{2}+\mu_{1}},\label{a3}
\end{equation}
\begin{equation}
a_{2}=\frac{a_{3}}{2}\left(\varphi_{4}-2\varphi_{3}+\varphi_{2}\right)+2\varphi_{4}-3\varphi_{3}+\varphi_{2},\label{a2}
\end{equation}
\begin{equation}
a_{4}=\frac{a_{6}+a_{7}}{2},\label{a4}
\end{equation}
\begin{equation}
a_{5}=\frac{a_{6}-a_{7}}{2},\label{a5}
\end{equation} with
\begin{equation}
a_{6}=\frac{a_{3}\left(\varphi_{2}-\mu_{1}\right)+2\varphi_{2}-\mu_{1}}{a_{2}}-3, \label{a6}
\end{equation}
\begin{equation}
a_{7}=\sqrt{\left(\frac{a_{3}\left(\varphi_{2}-\mu_{1}\right)+2\varphi_{2}-\mu_{1}}{a_{2}}-1\right)^{2}-4\frac{\mu_{1}(a_{3}+1)}{a_{2}}},
\end{equation}and
\begin{equation}\label{phi}
\varphi_{i}=\frac{\mu_{j}}{\mu_{j-1}},j > 1.
\end{equation}Above, $\mu_j$ is the $j$-th moment of $B$ and $\Gamma(.)$ is the Gamma function defined in \cite[Eq. (6.1.1)]{Abramowitz}. 
\end{prop}
\begin{IEEEproof}
 The RV $B$, given in \eqref{B}, is written as a sum of the RVs $|h_i| |g_{i}|$. Therefore, its PDF can be formulated in a closed-form approximation using the moment-based density approximants method presented in \cite{8245828}. The evaluation of the first four moments, ensuring an accurate approximation for the PDF of $B$, are obtained in \cite{LinaB} and are presented here for clarity and completeness of the work. Specifically, the first, second, third, and fourth moments are expressed as \eqref{mu_11}, \eqref{mu_22},  \eqref{mu_333}, and  \eqref{mu_444}, respectively. This completes the proof.

\begin{align}\label{mu_11}
\mu_{1} & =\frac{N\pi}{2},
\end{align}
\begin{align}\label{mu_22}
\mu_{2} =\left(4+(N-1)\frac{\pi^{2}}{4}\right)N,
\end{align}

\begin{equation}\label{mu_333}
\mu_{3}=\left\{ \begin{array}{l}
N\pi\left(\frac{9}{2}+6(N-1)+(N-1)(N-2)\frac{\pi^{2}}{8}\right),N\geq3\\
9\pi+3\times2\times\frac{\pi}{2}\times4=21\pi,N=2\\
\frac{9\pi}{2},N=1.
\end{array}\right.
\end{equation}
\end{IEEEproof}
\setcounter{equation}{16}
Based on the PDF obtained in \eqref{ccdf1}, an accurate approximation of the CDF of $B$ can be analytically computed as \cite{8245828}
\begin{equation}
F_{B}(x)\approx a_{1}a_2 G_{2,3}^{2,1}\left[\frac{x}{a_{2}}\left\vert \begin{array}{c}
1;a_{3}+1\\
a_{4}+1,a_{5}+1;0
\end{array}\right.\right], x\geq0 \label{ccdf12}.
\end{equation}

\subsection{Statistical Characterization of the Received Power}
\par \noindent Capitalizing on the statistical model of the E2E channel gain presented earlier, we derive the distribution of the instantaneous total received power at ER in the following proposition.
\begin{prop}
For an RIS-assisted WPT system, the CDF of the instantaneous power received at the ER node can be expressed as 
\begin{align}
F_{{P}_{r}}(x)&\approx a_{1}a_2 G_{2,3}^{2,1}\left[\frac{1}{a_{2}}\sqrt{\frac{x}{\bar{P}_{r}}}\left\vert \begin{array}{c}
1;a_{3}+1\\
a_{4}+1,a_{5}+1;0
\end{array}\right.\right], x\geq0 \label{cdf3}.
\end{align}
\end{prop}
\begin{IEEEproof}
 The CDF of $P_{s}$ is given as
\begin{equation}\label{cdfpr}
    F_{{P}_{r}}(x)=\text{Pr}(P_{r} \leq x).
\end{equation}Substituting \eqref{Pr-D2} in \eqref{cdfpr} yields \begin{equation}
F_{{P}_{r}}(x)=\text{Pr}\left(B \leq \sqrt{\frac{x}{\bar{P}_{r} }}\right)    
\end{equation}or equivalently
\begin{equation}\label{cdfpr2}
F_{{P}_{r}}(x)=F_{B}\left( \sqrt{\frac{x}{\bar{P}_{r} }}\right),
\end{equation} where
\begin{equation}\label{snravg}
\bar{P}_{r}=\frac{P_r}{d_{1}^\delta d_{2}^\delta}
\end{equation}denotes the average power received at ER. By invoking the expression given in \eqref{ccdf12}, \eqref{cdf3} can be obtained, which concludes the proof.
\end{IEEEproof}
\vspace*{0.3cm}
\noindent Accordingly, the PDF of $P_{s}$ can be obtained by applying \cite[Eq. (07.34.20.0001.01)]{wolfram} to differentiate \eqref{cdf3}, i.e.,
\begin{equation}
f_{P_{r}}(x)=\frac{d F_{{P}_{r}}(x)}{dx},   
\end{equation} yielding
\begin{align}
f_{{P}_{r}}(x)&\approx \frac{a_{1}a_2}{2x} G_{1,2}^{2,0}\left[\frac{1}{a_{2}}\sqrt{\frac{x}{\bar{P}_{r}}}\left\vert \begin{array}{c}
-;a_{3}+1\\
a_{5}+1,a_{4}+1;-
\end{array}\right.\right], x\geq0 \label{pdfpr}.
\end{align} 
\section{Battery Recharging Time Statistical Models}\label{BRTstat}
In this section, the statistical models developed in Section~II are employed to derive analytical expressions for the statistical characterization of the instantaneous BRT, $T_r$ at the ER node. 
\par Based on \cite{7374745}, the BRT of the ER node is inversely proportional to its received power, $P_r$ and is defined as
\begin{equation}\label{Tr}
T_r=\frac{\alpha}{P_r},
\end{equation}where $P_r$ is given in \eqref{Pr-D2}. Moreover, $\alpha$ denotes the conversion coefficient, which is a function of the battery and the RFEH circuit parameters; this can be expressed as follows 
\begin{equation}
\alpha=\frac{C_b D_d V_b}{\eta},
\end{equation}where $C_b$ is the battery capacity, $D_d$ is the battery discharge depth, $V_b$ is the charging voltage, and $\eta$ is the RF to direct currect efficiency.
In this section, we will exploit the PDF and CDF of the instantaneous received power, derived in \eqref{pdfpr} and \eqref{cdf3}, respectively, to obtain the statistical distribution of the BRT considering RIS-assisted WPT.

\subsection{Probability Density Function (PDF) of BRT}
\par \noindent The following proposition returns a closed-form expression for the PDF of the BRT when WPT is completed through an RIS.
\begin{prop}
For RIS-assisted WPT systems, the PDF of the battery recharging time at ER node is given as
\begin{align}\label{pdfTr}
f_{{T}_{r}}(\tau)&\approx \frac{a_{1}a_2}{2\tau} G_{1,2}^{2,0}\left[\frac{1}{a_{2}}\sqrt{\frac{\alpha}{\bar{P}_{r} \tau}}\left\vert \begin{array}{c}
-;a_{3}+1\\
a_{5}+1,a_{4}+1;-
\end{array}\right.\right], \tau>0.
\end{align}
\end{prop}
\begin{IEEEproof}
Using \eqref{Tr} and with the help of the Jacobian transformation method \cite{papoulis}, the PDF of $T_r$ can be given as
\begin{equation}\label{jacobian}
f_{T_r}(\tau)=\frac{\alpha}{\tau^2}f_{P_{r}}\left(\frac{\alpha}{\tau}\right),
\end{equation}which, with the aid of \eqref{pdfpr}, can be expressed in a closed-form as \eqref{pdfTr}. This completes the proof.
\end{IEEEproof}
\vspace*{0.3cm}
\noindent It is worth noting that \eqref{pdfTr} is simple and incorporates the Meijer G-function, which is a standard built-in function in most of the well-known mathematical software packages, such as MATLAB, MAPLE, and MATHEMATICA, and can therefore, be efficiently evaluated.

\subsection{Cumulative Distribution Function (CDF) of BRT}
\par\noindent The CDF of the BRT is defined as the probability that the instantaneous BRT falls below a predetermined threshold, $\tau_{th}$, i.e., $F_{T_r}(\tau_{th})=P\left( \tau \leq \tau_{th} \right)$.
Taking into account \eqref{Tr}, it is straightforward to note that the relation between the CDFs of the received power and the BRT becomes
\begin{equation}\label{cdftr}
F_{T_r}(\tau_{th})\approx 1-F_{P_{r}}\left(\frac{\alpha}{\tau_{th}}\right).
\end{equation}Therefore, by substituting \eqref{cdf3} into \eqref{cdftr}, we obtain the CDF of the BRT in a closed-form as
\begin{align}
F_{{T}_{r}}(\tau_{th})&=1- a_{1}a_2 G_{2,3}^{2,1}\left[\frac{1}{a_{2}}\sqrt{\frac{\alpha}{\bar{P}_{r}\tau_{th}}}\left\vert \begin{array}{c}
1;a_{3}+1\\
a_{4}+1,a_{5}+1;0
\end{array}\right.\right],\nonumber \\
& \quad\quad\quad\quad\quad\quad\quad\quad\quad\quad\quad\quad \quad\quad\quad\quad \tau_{th}>0. \label{cdftr2}
\end{align}

\subsection{BRT Mean Value, Variance, Skewness, Kurtosis, and AoF} \label{meanBRT}
\par \noindent The $n$-th order moment of the BRT, denoted by $\mu_{Tr}(n)$, is a very useful statistical tool, as it enables the characterization of the mean value of the BRT, in addition to other underlying useful properties such as its skewness and kurtosis. Moreover, it can be employed to quantify the AoF, as will be elaborated next.

\newcounter{tempequationcounter3}
\begin{figure*}[ht]
\normalsize
\setcounter{equation}{39}
\begin{align}\label{BRTCLT}
f_{\tau_r}(\tau)= \frac{2\alpha}{\tau^2 N\left(16-\pi^2\right)\bar{P}_{r}}\left(\frac{\tau N^2\pi^2 \bar{P}_{r}}{4\alpha}\right)^{\frac{1}{4}}\text{exp}\left(-\frac{(4\alpha/\tau)+N^2\pi^2\bar{P}_{r}}{2N\left(16-\pi^2\right)\bar{P}_{r}}\right)\textit{I}_{-\frac{1}{2}}\left(\frac{2\pi}{\left(16-\pi^2\right) \bar{P}_{r}}\sqrt{\frac{\bar{P}_{r} \alpha}{\tau}}\right), \tau>0.
\end{align}
\hrulefill
\end{figure*}
\setcounter{equation}{30}
\par Having \eqref{pdfTr} in hand, one can derive the $n$-th moment of the BRT through the $n$-th order statistical expectation, as presented in the following proposition.
\begin{prop}
The $n$-th moment of the BRT of an RIS-assisted WPT system can be expressed in a simple closed-form as
\begin{align}\label{momentTr}
\mu_{T_r}(n)\approx a_1 a_2^{(1-2n)}\left(\frac{\alpha}{\bar{P}_{r}}\right)^{n} \frac{\Gamma(a_4+1-2n)\Gamma(a_5+1-2n)}{\Gamma(a_3+1-2n)}.
\end{align}
\end{prop}
\begin{IEEEproof}
The $n$-th moment of $T_r$ can be evaluated in a straightforward manner by taking the statistical expectation 
\begin{equation}\label{moment}
\mu_{T_r}(n)=\int_0^\infty \tau^n f_{{T}_{r}}(\tau) d\tau.    
\end{equation}Then, by substituting \eqref{pdfTr} in \eqref{moment}, we obtain
\begin{equation}\label{moment2}
\mu_{T_r}(n)=\frac{a_1 a_2}{2}\mathcal{J}
\end{equation} where
\begin{equation}
\mathcal{J}=\int_0^\infty \tau^{n-1} G_{1,2}^{2,0}\left[\frac{1}{a_{2}}\sqrt{\frac{\alpha}{\bar{P}_{r} \tau}}\left\vert \begin{array}{c}
-;a_{3}+1\\
a_{5}+1,a_{4}+1;-
\end{array}\right.\right]d\tau.    
\end{equation} By applying \cite[Eq. (07.34.21.0009.01)]{wolfram} to solve the integral, $\mathcal{J}$, in a closed-form, the expression in \eqref{momentTr} can be obtained. This completes the proof.    
\end{IEEEproof}
\vspace*{0.3cm}
The expression in \eqref{momentTr} can be used to obtain the mean value of the BRT, $\bar{T}_r$ by setting $n=1$, i.e., $\bar{T}_r=\mu_{T_r}(1)$. Also, the variance, $\sigma^2_{\tau_r}$, can be computed as 
\begin{equation}
\sigma^2_{\tau_r}=\mu_{T_r}(2)-\bar{T}_r^2.  
\end{equation} 
The AoF parameter is viewed as a unified statistical measure of the severity of fading. Therefore, it is useful in quantifying the robustness of the RIS-assisted WPT technology against channel fading. The AoF is defined in \cite[Eq.(1.27)]{Simon} as the ratio of the variance to the square mean of the instantaneous received power. Subsequently, by employing \eqref{momentTr}, $AoF=\sigma^2_{\tau_r}/\bar{T}_r^2$.
\par \textit{Remark: By examining \eqref{momentTr}, when $n$ is set to 1, we note that the scaling law of the mean value of the BRT as a function of the RIS distances from the source and the ER nodes, respectively denoted as $d_1$ and $d_2$, and which are implicitly defined through $\bar{P}_r$ in \eqref{snravg}, dictates that $\bar{T}_r$ increases with the square of the product of the distances $d_1$ and $d_2$. This suggests that the minimum value of $\bar{T}_r$ is achieved when RIS is located either closer to the source or the ER node.}
\newcounter{tempequationcounter1}
\begin{figure*}[!t]
\normalsize
\setcounter{equation}{44}
\begin{equation}\label{snrCLT}
f_{P_{r}}(x)= \frac{2}{N\left(16-\pi^2\right)\bar{P}_{r}}\left(\frac{N^2\pi^2 \bar{P}_{r}}{4x}\right)^{\frac{1}{4}}\text{exp}\left(-\frac{4x+N^2\pi^2\bar{P}_{r}}{2N\left(16-\pi^2\right)\bar{P}_{r}}\right)\textit{I}_{-\frac{1}{2}}\left(\frac{2\pi}{\left(16-\pi^2\right) \bar{P}_{r}}\sqrt{\bar{P}_{r} x}\right), x\geq0 
\end{equation}
\hrulefill
\end{figure*}
\setcounter{equation}{35}
\par In addition to the mean and variance of the BRT, statistical properties such as skewness, denoted by $\epsilon $, and kurtosis, denoted by $\Psi$, can be evaluated from \eqref{momentTr} to provide a deeper insight of the distribution of the BRT. More specifically, the skewness measures the asymmetry of the PDF of the BRT about its mean value, while the kurtosis is an indicator of its peakedness or flatness and the heaviness of its tail. The skewness can be expressed as \cite{papoulis}
\begin{equation}
\epsilon =\frac{\mu_{T_r}(3)}{\mu^{3/2}_{T_r}(2)},
\end{equation}while the kurtosis can be given as
\begin{align}
\Psi &=\frac{\mu_{T_r}(4)}{\mu^2_{T_r}(2)}-3.
\end{align}
\noindent\underline{Special Case}: In the special case when RIS consists of one RE $(N=1)$, $B=\left|h\right|\left|g\right|$ is modeled as a double Rayleigh distribution. Accordingly, the PDF of the total received power at ER is given as \cite{Simon} 
\begin{equation}
f_{P_{r}}\left(x\right)=\frac{2}{\bar{P}_{r}}K_{0}\left(2\sqrt{\frac{x}{\bar{P}_r}}\right),x\geq 0.\label{PDFN1}
\end{equation}By substituting \eqref{PDFN1} in \eqref{jacobian}, the PDF of the BRT can be expressed as 
\begin{equation}
f_{\tau_r}\left(\tau\right)=\frac{2\alpha}{\bar{P}_{r} \tau^2}K_{0}\left(2\sqrt{\frac{ \alpha}{\bar{P}_{r}\tau}}\right),\tau>0.\label{PDFTauN1}
\end{equation}
\setcounter{equation}{40}
\noindent\underline{Special Case:} In the special case when the RIS consists of an asymptotically large number of REs $(N\gg 1)$, the following lemma returns a closed-form expression for the PDF of the BRT for WPT systems.
\begin{lem}
For a sufficiently large number of RIS REs, $N\gg 1$, the PDF of the BRT for RIS-assisted WPT system can be expressed in a closed-form as \eqref{BRTCLT}, shown at the top of this page.
\end{lem}
\begin{IEEEproof}
Recalling that $|h_i|$ and $|g_i|$ are independently Rayleigh distributed RVs, then the mean and variance of $B_i$ are, respectively, given as 
\begin{equation}
 \mathbb{E}[B_i]=\mathbb{E}[|h_i||g_i|]=\frac{\pi}{2}   
\end{equation} and 
\begin{equation}
\text{VAR}[B_i]=\left(\frac{16-\pi^2}{4}\right).
\end{equation} As the number of RIS elements becomes sufficiently large, then according to the CLT $B$ converges to a Gaussian distribution with mean
\begin{equation}
\mu_{N}=\mathbb{E}[B]=\frac{N\pi}{2}
\end{equation} and variance 
\begin{equation}
 \sigma_N^2=N\left(\frac{16-\pi^2}{4}\right),   
\end{equation} where $\mu_{N}$ and $\sigma_N^2$ can be deduced by using \eqref{mu_11} and \eqref{mu_22}. As a result, the instantaneous received power at ER, $P_{s}$, defined in \eqref{Pr-D2}, presents a non-central chi-square distribution with one degree of freedom \cite{Proakis}, and its PDF can be expressed as \eqref{snrCLT}, shown at the top of this page, where $\textit{I}_v(.)$ is the modified Bessel function of order $v$ defined in \cite[Eq. (9.6.20)]{Abramowitz}. Finally, The expression in \eqref{BRTCLT} can be obtained by employing \eqref{snrCLT} into \eqref{Tr}. This completes the proof. 
\end{IEEEproof}
\par Based on Lemma 1 and using \eqref{BRTCLT}, the mean value of the BRT when $N\gg 1$ converges to
\setcounter{equation}{45}
\begin{equation}\label{meanclt}
\bar{T}_r=\mathbb{E}[\tau_r]= \frac{4\alpha}{\left(N^2 \pi^2 + N \left(16-\pi^2\right)\right)\bar{P}_{r}}.   
\end{equation}By direct inspection of \eqref{meanclt}, we note that the mean value of the BRT in RIS-assisted WPT systems is shown to be inversely proportional to the square of the total number of REs, $N$, i.e., $\mathbb{E}[\tau_r]\propto \frac{1}{N^2 \bar{P}_{r}}$. This result is in agreement with the BRT definition given in \eqref{Tr}.

\section{Numerical and Simulation Results}\label{results}
In this section, numerical and Monte Carlo simulation results are presented to validate the accuracy of the proposed theoretical framework. This section also focuses on characterizing the properties of the BRT in RIS-assisted WPT wireless systems. The term Monte Carlo simulations refers to the use of actual fading channel variates with a number of repetitions of $10^6$ trials. 
\par Unless otherwise stated, the RFEH efficiency factor $\eta$~=~0.5, as a worst case scenario, capturing the effects of low-cost hardware, and the total distance, $d_{tot}$, between the source node, $S$, and the ER node is set to 5~m. In order to ensure far-field WPT, we assume that the size of RIS is relatively smaller than the transmission distance. It is assumed that the RIS is located mid-way between $S$ and ER, i.e., $d_1=d_2=d_{tot}/2$, and the path-loss exponent, $\delta=2.7$ \cite{8539989}. Also, it is recalled that $P_s$ defines the total transmit power of the system. All simulation parameters, including the ER battery parameters, are summarized in Table 1.
\begin{table}[h]
\centering
\caption{Simulation Parameters.}
\label{t2}
\begin{tabular}{|l|c|c|}
\hline
\multicolumn{1}{|c|}{\textbf{Name}} & \textbf{Symbol} & \textbf{Value}               \\ \hline
RFEH efficiency of the ER circuit      & $\eta$            & 0.5                          \\ \hline
Normalized $S\to$RIS distance      & $d_{1}/d_{tot}$      & 0.5                          \\ \hline
Normalized RIS$\to$ER distance      & $d_{2}/d_{tot}$      & 0.5                          \\ \hline
Path-loss exponent                  & $\delta$       & 2.7                          \\ \hline
Battery capacity                    &$C_b$           & 10~mAh
           \\ \hline
Discharge depth                    &$D_d$           & 0.4
           \\ \hline
Battery charging voltage            &$V_b$           & 1.2~V
           \\ \hline
\end{tabular}
\end{table}
\begin{figure}[!t]
\centering
\includegraphics[width=3.5in]{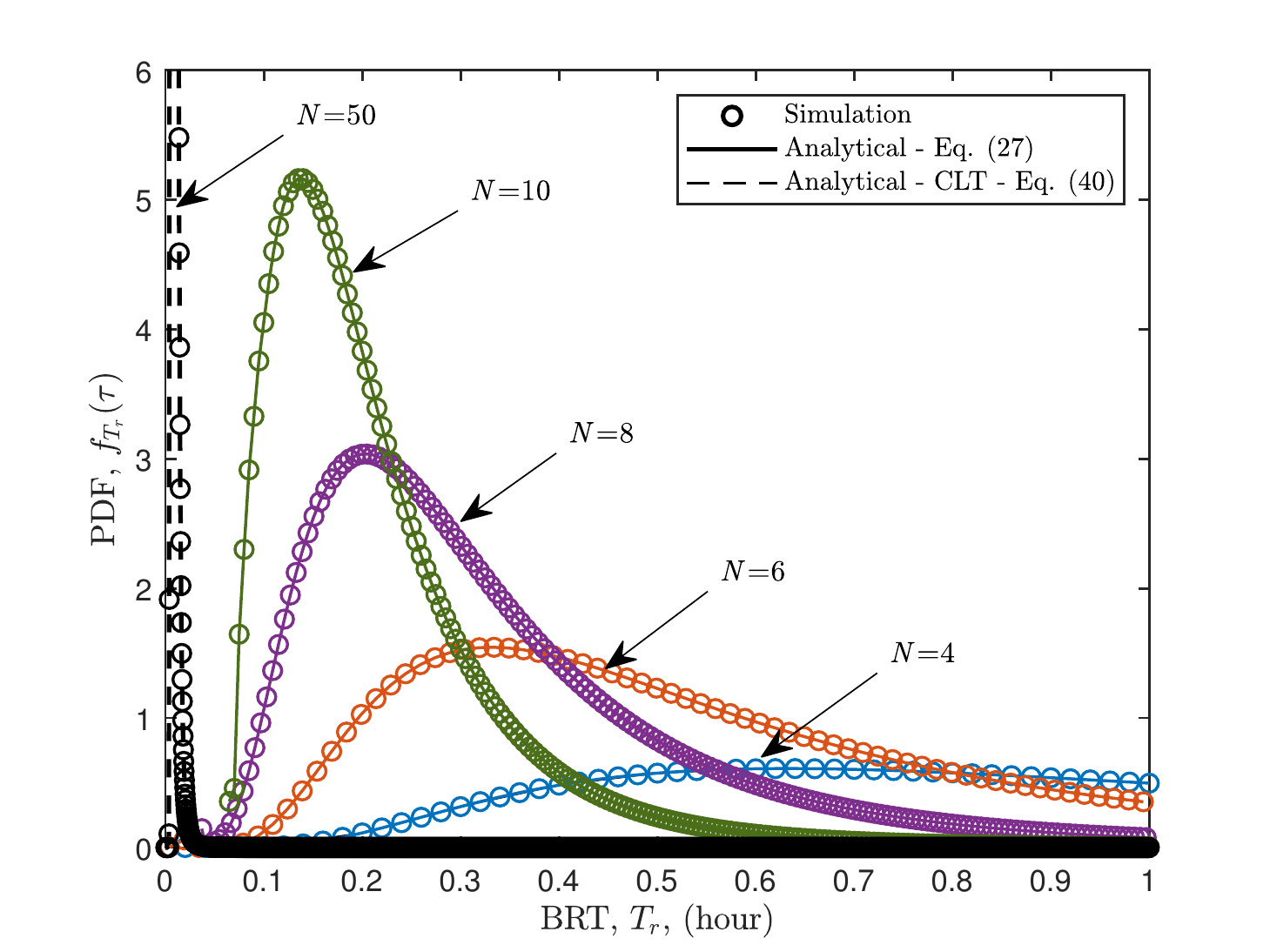} 
\caption{The PDF of the BRT of an RIS-assisted WPT system, for different values of $N$, where the transmit power, $P_s$~=~15~dBm. The PDFs obtained through the CLT are also plotted for $N=$ 10 and 50.}
\label{fig1} 
\end{figure}
\par In Fig.~\ref{fig1}, the PDF of the BRT for RIS-assisted WPT systems is illustrated for $N$ RIS elements, using the analytical expression given in \eqref{pdfTr}, where $N=$ 4, 6, 8, 10, 50. Additionally, the PDF of the BRT obtained through the CLT, as given in \eqref{BRTCLT}, is also shown for $N=$ 10 and 50. It is observed that the simulation results are in full agreement with the analytical PDF curves, reflecting the accuracy of our proposed mathematical model and its effectiveness in capturing the statistical properties of the BRT. Additionally, we notice that as $N$ increases, the BRT value of the ER node decreases. In more details, the highest probable value of the BRT drops from 0.64 to 0.2~hr (38.4 to 12~mins) when the number of REs increases from 4 to 8. Finally, it is evident that as the number of RIS elements increases through the CLT, the PDF converges to an impulse function, indicating that increasing $N$ causes a significant reduction in the BRT of WPT systems, due to the enhanced spatial diversity gain of RIS-assisted systems, and thus, making them particularly attractive for large-scale RFEH applications.
It is worth mentioning that the effect of increasing the number of RF sources on the BRT was studied in \cite{7374745}. Although their results demonstrated that the BRT decreases notably by increasing the number of RF sources, this comes at the cost of extra transmission power needed for each additional RF source. 
\begin{figure}[!t]
\centering
   \begin{subfigure}[b]{0.5\textwidth}
   \includegraphics[width=3.5in]{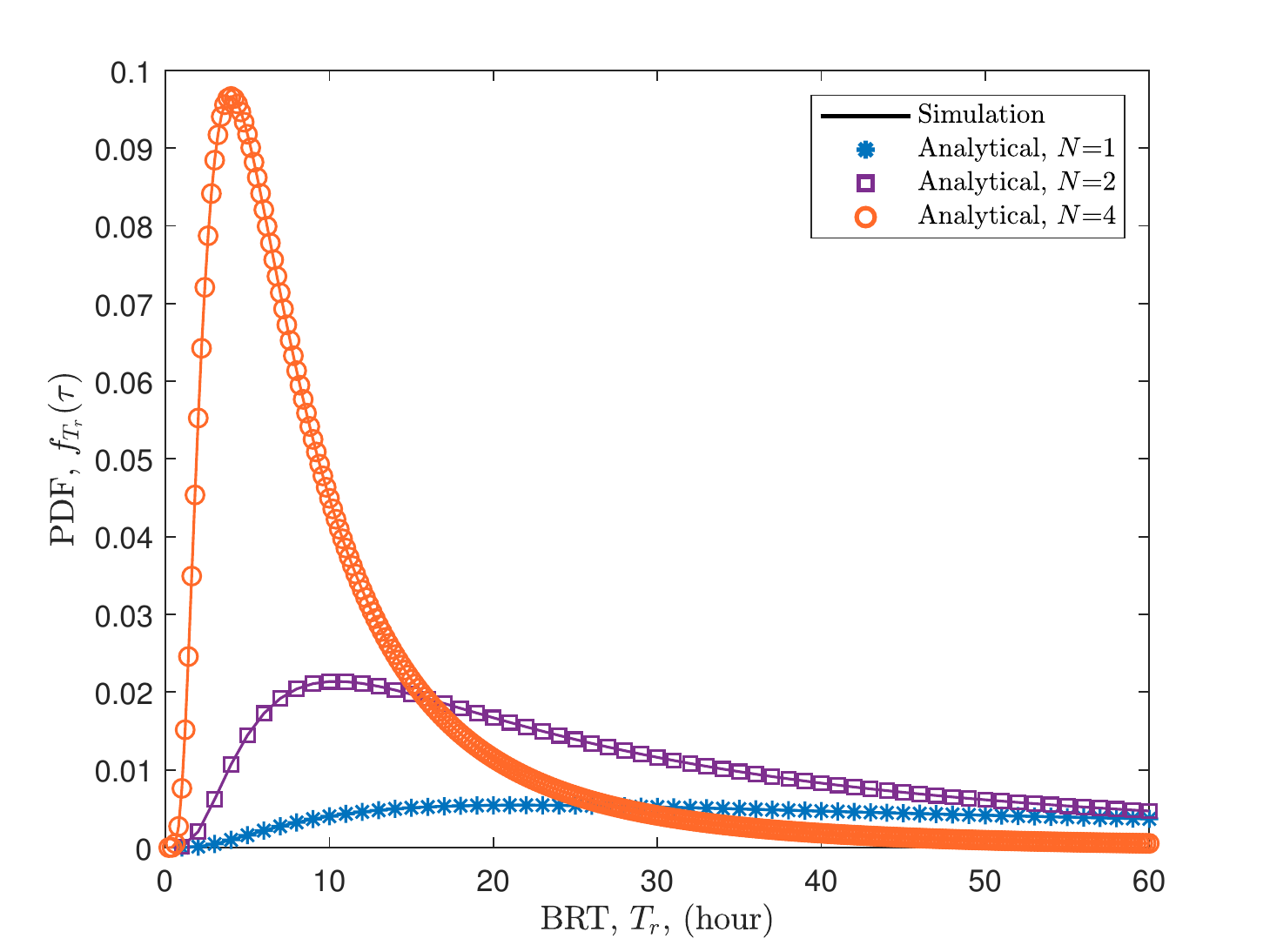}
\caption{$P_s$~=~7~dBm}
   \label{snr7} 
\end{subfigure}\hfill
\begin{subfigure}[b]{0.5\textwidth}
   \includegraphics[width=3.5in]{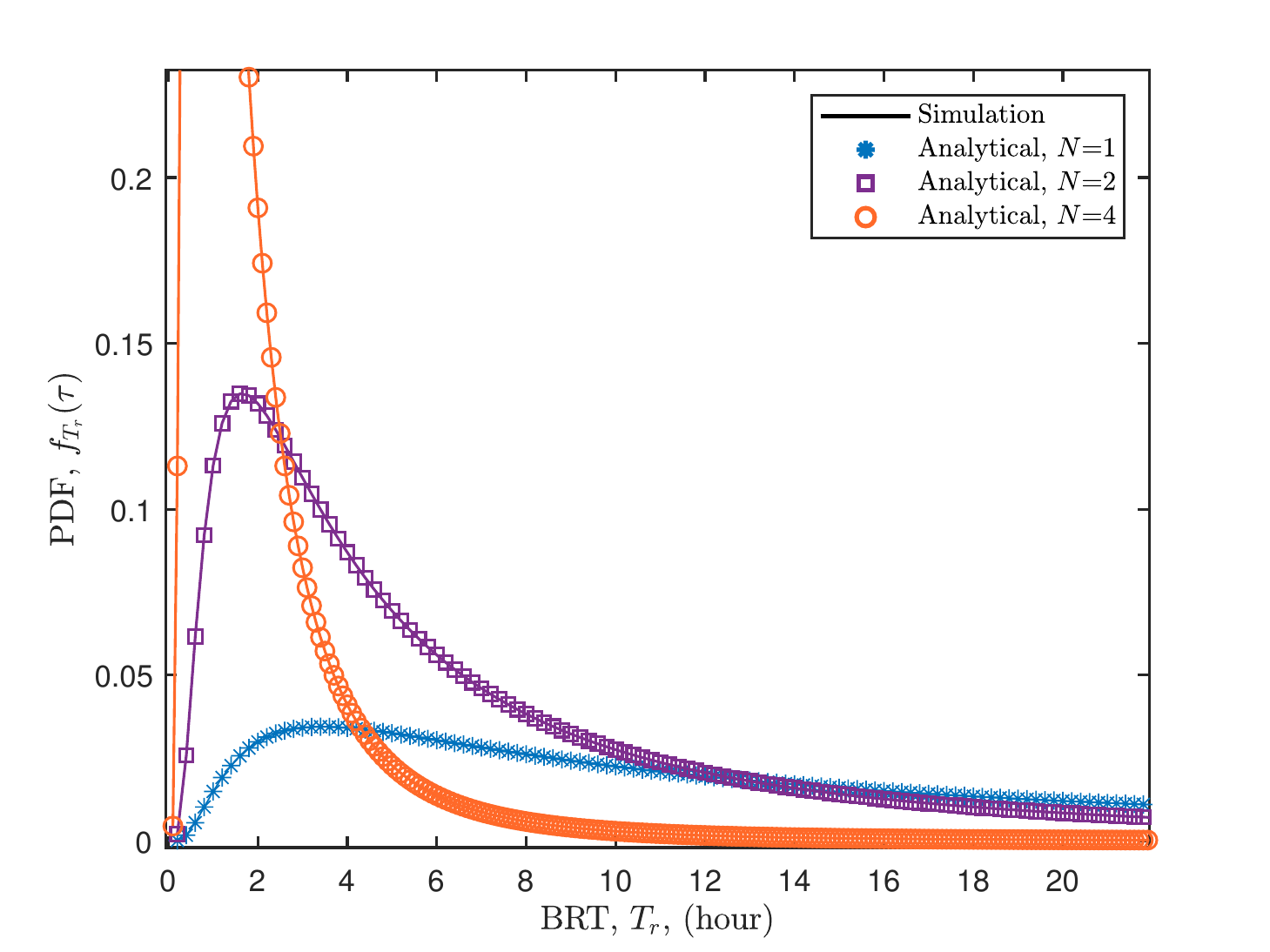}
   \caption{$P_s$~=~15~dBm}
   \label{snr15}
\end{subfigure}\hfill
\begin{subfigure}[b]{0.5\textwidth}
   \includegraphics[width=3.5in]{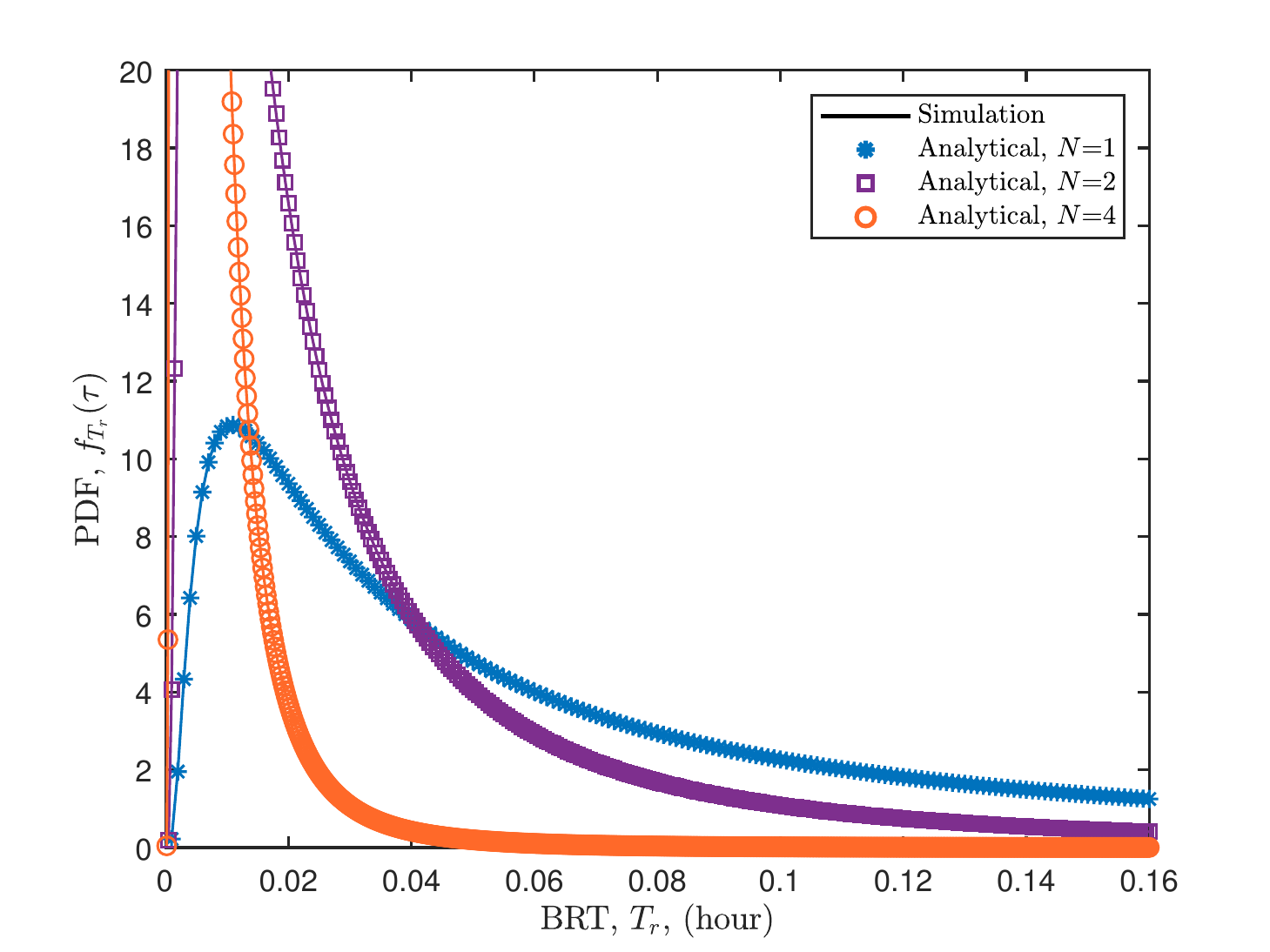}
   \caption{$P_s$~=~40~dBm}
   \label{snr40}
\end{subfigure}\hfill
\caption{The PDF of the BRT in RIS-assisted WPT systems, in (a) low transmit power, $P_s$~=~7~dBm, (b) moderate transmit power, $P_s$~=~15~dBm, and (c) high transmit power, $P_s$~=~40~dBm, regimes, and for different values of $N$.}
 \vspace*{-0.5cm}
 \label{fig2} 
 \end{figure}
 \par In Fig.~\ref{fig2}, we investigate the effect of varying the value of the transmit power, $P_s$, on the BRT performance of RIS-assisted WPT. We plot the PDF of the BRT when $N=1,2$, and 4, using the analytical expressions obtained in \eqref{pdfTr} and \eqref{PDFTauN1}. Figs.~\eqref{snr7}, \eqref{snr15}, and \eqref{snr40} present, respectively, low, moderate, and high transmit power regimes, i.e., $P_s=$7, 15, and 40~dBm. By closely inspecting these figures, it is clearly observed that when higher transmit power values are encountered, the variance of the distribution of the BRT decreases while its kurtosis increases (i.e., sharper PDF peak), indicating the improvement in the BRT predictability. Interestingly, it can be also noted that even in the low transmit power scenario, depicted in Fig.~\ref{snr7}, the mean value of the BRT exhibits high predictability when $N=4$ compared to the cases when $N=1$ or 2. This indicates that the deployment of an RIS is significantly promising in enhancing the reliability of the WPT process, and may be attained by adding low-cost passive RIS elements (featuring no transmit power consumption) instead of increasing the source transmission power. Further details about this aspect will be provided in subsequent discussions.  
\par
\begin{figure}[!t]
\centering
\includegraphics[width=3.5in]{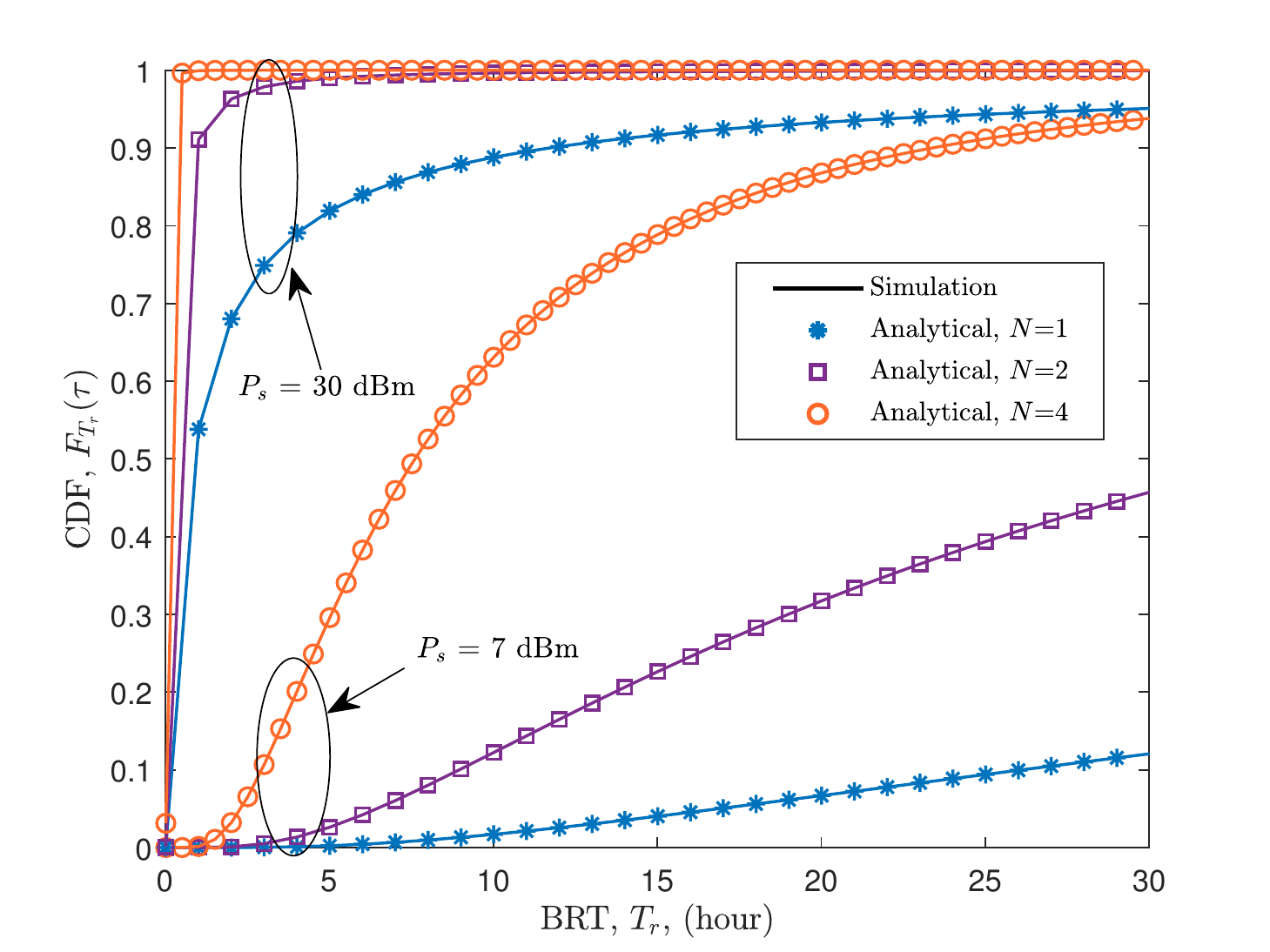} 
\caption{The CDF as a function of BRT threshold for RIS-assisted for low- and high- transmit power scenarios and for different values of $N$. }
\label{fig3} 
\vspace{-.3cm}
\end{figure}

 To gain more insights about the effect of varying the transmit power on the statistical distribution of the BRT in RIS-assisted WPT systems, we provide in Fig.~\ref{fig3} the CDF as a function of the BRT threshold, $\tau_{th}$, of the RIS-assisted system using the expression in \eqref{cdftr}. The examination is carried out for different $N$ values and assuming two transmit power scenarios, namely low- ($P_s=$~7~dBm) and high- ($P_s=$~30~dBm) transmit power. The excellent fit between the simulation and the analytical results verify the accuracy of our developed theoretical framework. As expected, for a fixed $N$, as the $\tau_{th}$ increases, the CDF value increases. For example, for $N=4$ and $P_s$~=~7~dBm, as $\tau_{th}$ changes from 5 to 10~hrs, the CDF value approximately doubles. Additionally, for a given $\tau_{th}$ value, as $N$ increases, the CDF value improves. This indicates that the efficiency of the RFEH process is remarkably improved in an RIS-assisted WPT system by increasing the number of REs, $N$. For instance, at $P_s$=~30~dBm, $\tau_{th}$ can be reduced by about 13 hrs by employing an RIS with 2 elements instead of 1 in order to achieve a targeted probability of charging of 0.9. Fig.~\ref{fig3} also demonstrates that the steepness of the CDF curve increases as the value of the transmit power shifts from low to high. This indicates that adding more RIS elements is more rewarding in the low transmit power than the high transmit power case, where the reduction in the BRT value is more significant. 
\par
\begin{figure}[!t]
\centering
\includegraphics[width=3.5in]{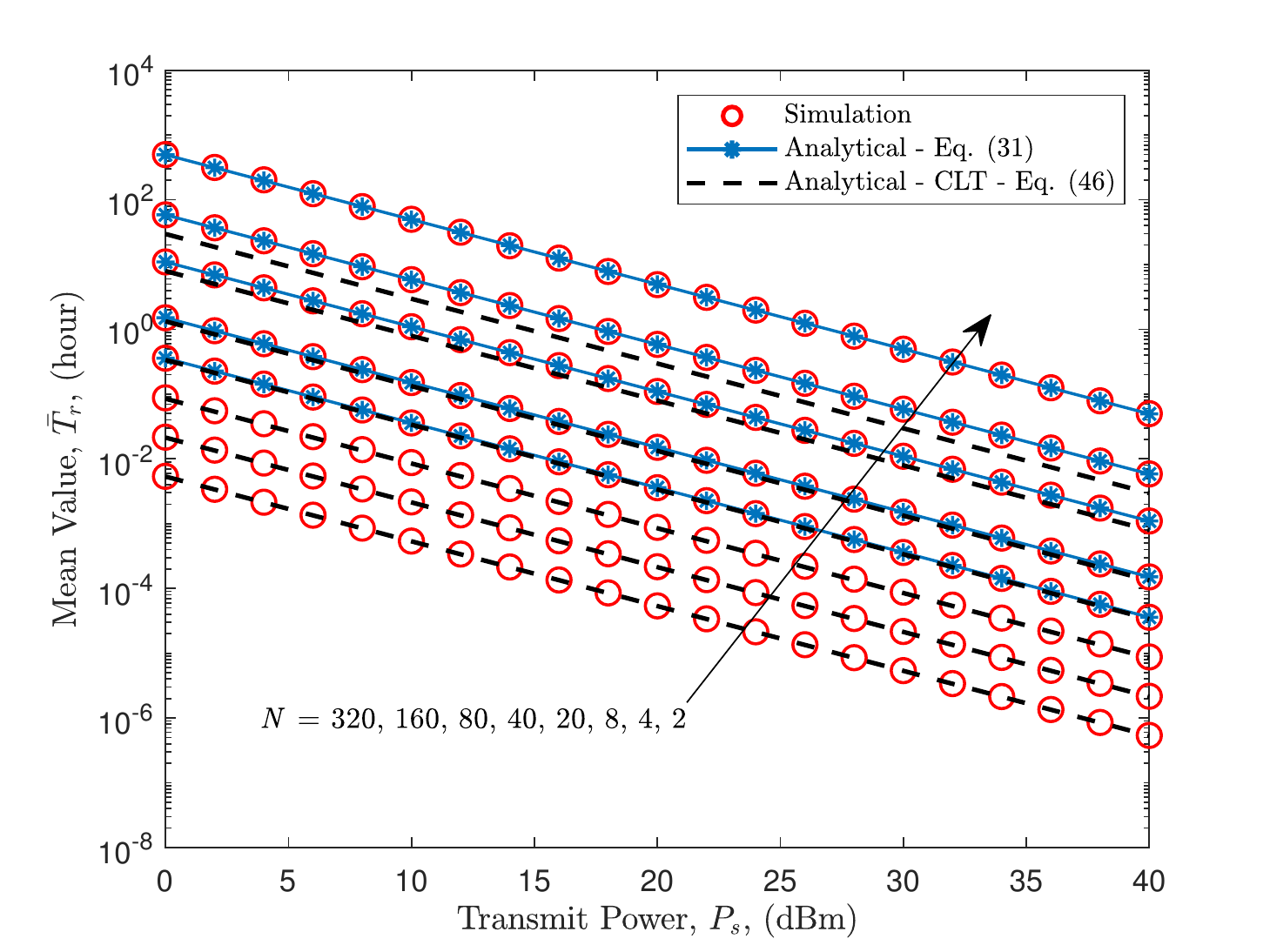} 
\caption{Mean value of the BRT as a function of the transmit power in RIS-assisted WPT systems, for different values of $N$.}
\label{fig4} 
\end{figure}

 Fig.~\ref{fig4} depicts the mean value of the BRT, $\bar{T}_r$, as a function of the total transmit power, $P_s$, for different values of $N$. We also examine the convergence of the mean value, obtained analytically in \eqref{momentTr}, towards that obtained through the CLT, given in \eqref{meanclt}. It can be deduced from Fig.~\ref{fig4} that $\bar{T}_r$ converges to that computed via the CLT even at relatively small values of $N$; for example, when $N>8$. This suggests that in such setups, the mean value of the BRT in RIS-assisted WPT systems can be mathematically evaluated using the simpler expressions in \eqref{meanclt} instead of \eqref{momentTr}, which is computationally more demanding, since it involves the evaluation of the parameters $a_1$ through $a_5$. We further notice that as $N$ and $P_s$ increase, the mean value of the BRT linearly decreases. This finding is in agreement with the theoretical result obtained in \eqref{meanclt}. In more details, for a given value of $P_s$, doubling the number of deployed RIS elements, $N$, yields a reduction in the mean value of the BRT by about 4 times. In summary, the choice of $N$ depends on whether the application scenario operates in the low or high transmit power scenarios.
\par \textit{Remark: We emphasize that the analytical expressions derived in \eqref{momentTr} and \eqref{meanclt} lend themselves as effective tools in determining the minimum number of RIS elements that should be deployed in order to achieve a feasible BRT value for a targeted energy efficiency in WPT systems, while avoiding the need of unnecessary phase adjustments to account for the extra deployed RIS elements.}
\begin{figure}[!t]
\centering
\includegraphics[width=3.5in]{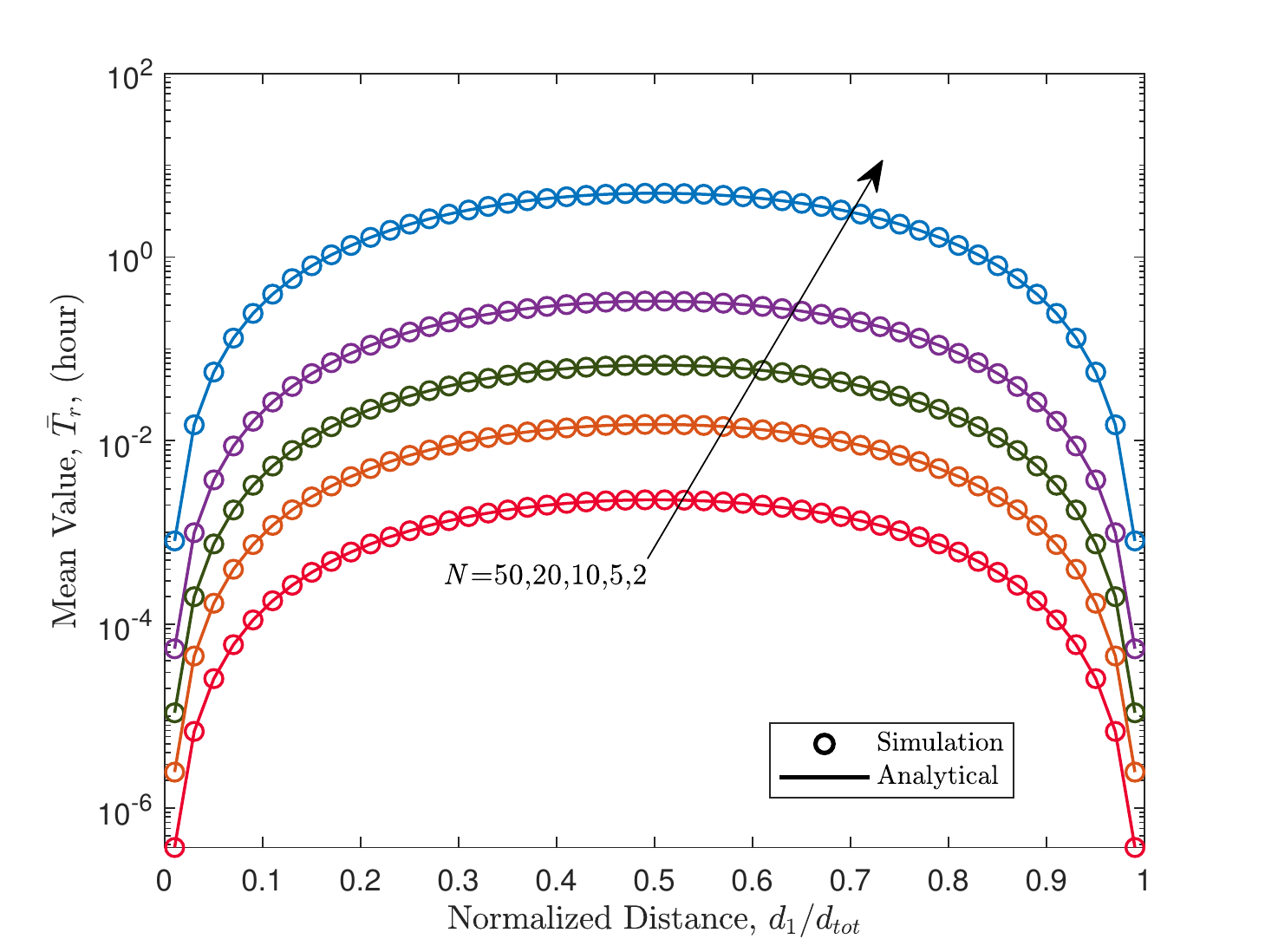} 
\caption{Mean value of the BRT as a function of the normalized $S\to$RIS distance, $d_1/d_{tot}$, for RIS-assisted WPT systems, for different values of $N$ and $P_s$~=~20~dBm. }
\label{fig5} 
\end{figure}
\par To quantify the impact of the RIS location on the BRT performance, in Fig.~\ref{fig5}, we inspect the behavior of the mean value of the BRT, $\bar{T}_r$, as a function of the normalized $S\to$RIS distance, $d_1/d_{tot}$, for different values of $N$ and fixed $P_s$=20~dBm. We set the normalized RIS$\to$ER distance to $d_2/d_{tot}=1-d_1/d_{tot}$. It is noted that for a given RIS location, the energy efficiency of a WPT system can be enhanced by equipping the RIS with more REs. This key BRT performance insight is indeed beneficial when there is no flexibility in choosing the location of the RIS due to the layout or geometry of the application environment. Additionally, it is straightforwardly observed that the minimum mean value of BRT is attained when the RIS is located either closer to $S$ or the ER node. This verifies our theoretically proven result, as discussed in Sec.~\ref{meanBRT} following \eqref{momentTr}. Therefore, it is not surprising to note that adding RIS elements is more rewarding, with respect to $\bar{T}_r$, when RIS is located mid-way between the $S$ and ER nodes.
\par \textit{Remark: The investigation carried out above would enable system design engineers to extract fundamental performance insights to choose the optimal placement of the RIS and/or IoE nodes in large-scale WPT systems.}
\par
\begin{figure}[!t]
\centering
\includegraphics[width=3.5in]{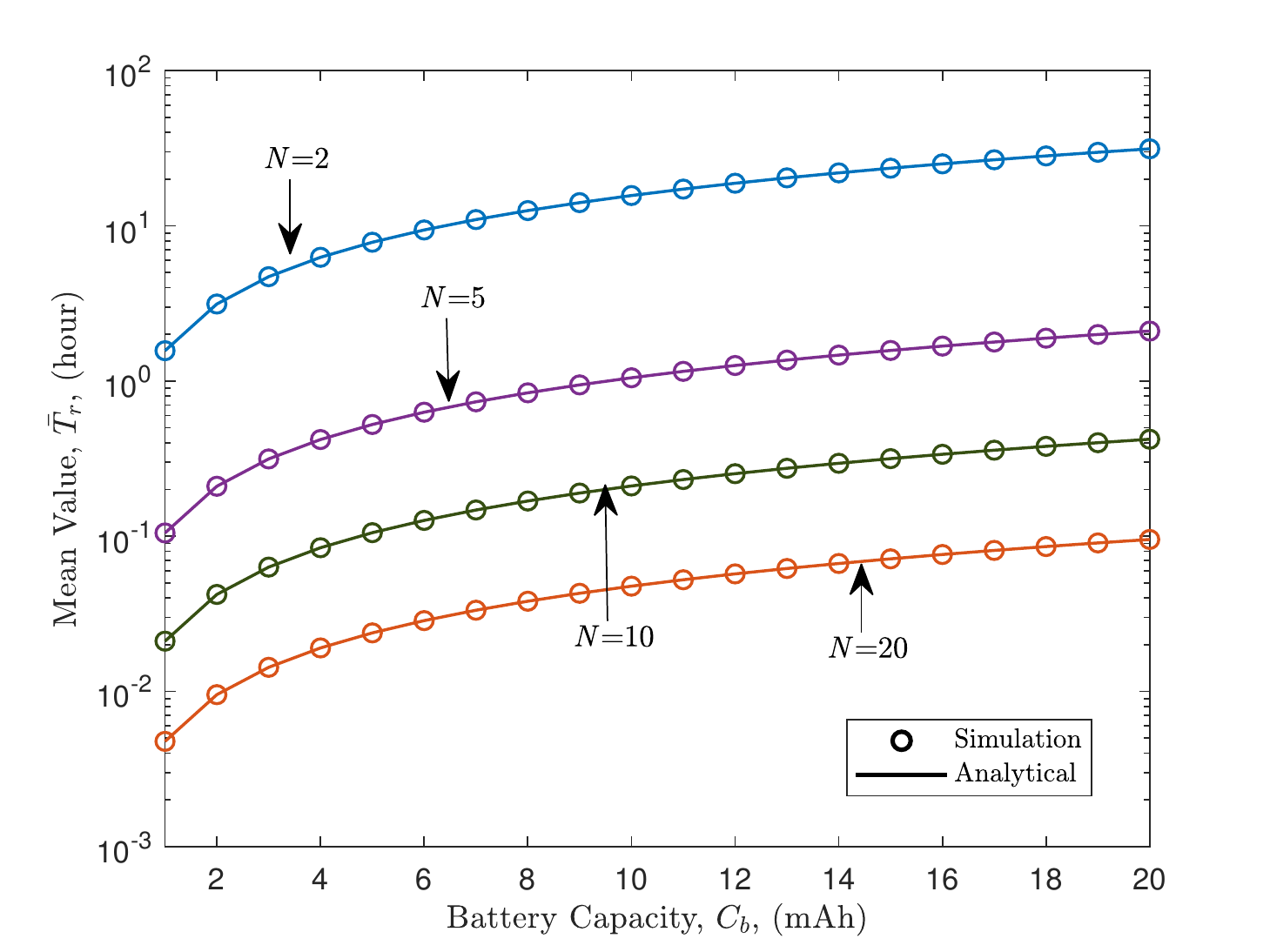} 
\caption{Mean value of the BRT as a function of the battery capacity of the ER node in RIS-assisted WPT systems, for different values of $N$ and when $P_s$~=~20~dBm.}
\label{fig6} 
\end{figure}

 To address the effect of varying the battery parameters, in Fig.~\ref{fig6} we plot the mean value of the BRT, $\bar{T}_r$, with respect to the battery capacity, $C_b$, of the ER node when the value of the transmit power is fixed to 20~dBm. As expected, for a given $N$ value, it is observed that as $C_b$ increases, the mean value of the BRT increases as well. However, one can take the advantage of adding RIS elements to dramatically boost the sustainability of WPT systems without changing the battery parameters of the RFEH device. In other words, this figure reveals that, independent of $C_b$, as $N$ doubles in value, $\bar{T}_r$ decreases by about 5 times. For example, for a given $C_b=10$~mAh, as $N$ is varied from 5 to 10 to 20, the mean value of the BRT is reduced from 1~hr to 12.65~mins to 2.86~mins. Therefore, it turns out that our proposed mathematical tools are useful in identifying the design parameters of IoE devices for future RIS-assisted WPT applications. This can be of considerable advantage for applications where the battery capacity implies constraints on the shape and size of the RFEH nodes, such as wearables or implantables. 
\begin{figure}[!t]
\centering
\includegraphics[width=3.5in]{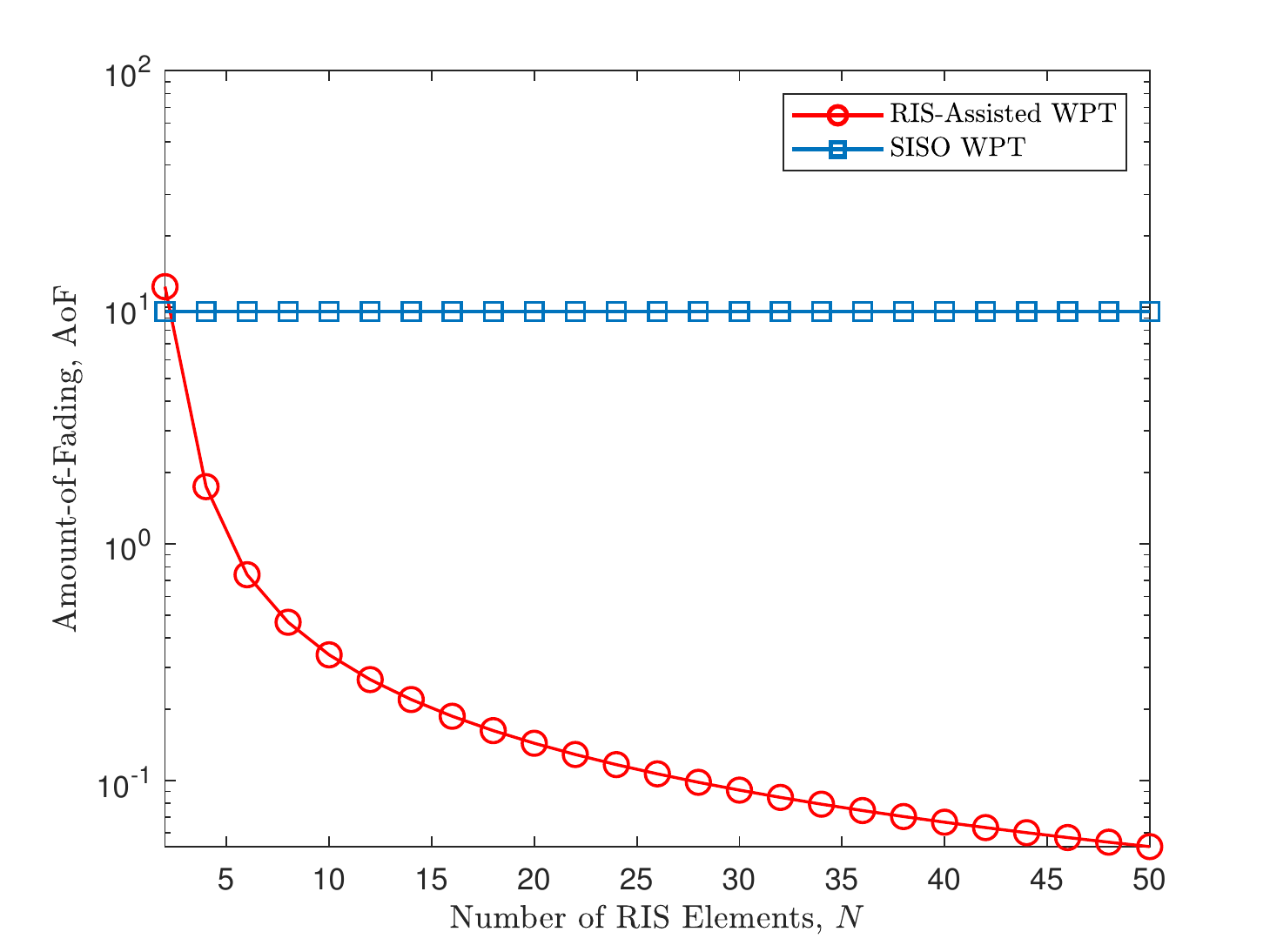} 
\caption{AoF as a function of $N$ when $P_s$~=~20~dBm.}
\label{fig7} 
\end{figure}
\par Finally, in Fig.~\ref{fig7}, we examine the AoF for RIS-assisted WPT as a function of the number of REs, $N$. We also plot the AoF for single-input single-output (SISO) WPT as a benchmark. Fig.~\ref{fig7} clearly shows that our analytical framework is useful in quantifying the robustness of RIS-assisted WPT to fading channels. Particularly, Fig.~\ref{fig7} reveals that increasing the number of REs reduces the AoF of the end-to-end channel, thereby improving the efficiency of WPT undergoing fading.
\section{Conclusion}\label{conc}
In this paper, we developed a theoretical framework to investigate the energy sustainability of RIS-assisted WPT systems, from the BRT perspective of an RFEH node. In particular, assuming that WPT is completed over Rayleigh fading channels, we provided the statistical characterization of the instantaneous received power of the system, including its PDF and CDF. Based on that, we derived novel low-complexity tight closed-form approximations for the PDF, CDF, and moments of the BRT as functions of the received power, battery parameters, and number of RIS REs. Additionally, using the CLT, we derived closed-form expressions for the PDF and mean value of the BRT considering that RIS is equipped with a large number of elements. Besides being accurate and mathematically tractable, our results reveal that the proposed statistical tools provide an efficient means to evaluate RIS-assisted WPT systems and extract useful design insights. For example, our results show that doubling the number of RIS elements improves the predictability of the BRT of the RFEH nodes in the network and offers a 4-fold reduction in its mean value. Moreover, it is reported that the characteristics of the BRT are significantly impacted not only by the system parameters, such as the distance between the nodes, but also by the battery parameters of the RFEH node, such as the battery capacity. Finally, our results illustrated that significant performance gains in the BRT have been observed by locating the RIS close to the source or to the RFEH node.

\balance 

\bibliographystyle{IEEEtran}
\bstctlcite{BSTcontrol}
\bibliography{RIS_ChargingTime_V1}

\begin{thebibliography}{10}
\providecommand{\url}[1]{#1}
\csname url@samestyle\endcsname
\providecommand{\newblock}{\relax}
\providecommand{\bibinfo}[2]{#2}
\providecommand{\BIBentrySTDinterwordspacing}{\spaceskip=0pt\relax}
\providecommand{\BIBentryALTinterwordstretchfactor}{4}
\providecommand{\BIBentryALTinterwordspacing}{\spaceskip=\fontdimen2\font plus
\BIBentryALTinterwordstretchfactor\fontdimen3\font minus
  \fontdimen4\font\relax}
\providecommand{\BIBforeignlanguage}[2]{{%
\expandafter\ifx\csname l@#1\endcsname\relax
\typeout{** WARNING: IEEEtran.bst: No hyphenation pattern has been}%
\typeout{** loaded for the language `#1'. Using the pattern for}%
\typeout{** the default language instead.}%
\else
\language=\csname l@#1\endcsname
\fi
#2}}
\providecommand{\BIBdecl}{\relax}
\BIBdecl

\bibitem{di2019smart}
M.~Di~Renzo \emph{et~al.}, ``Smart radio environments empowered by
  reconfigurable {AI} meta-surfaces: an idea whose time has come,''
  \emph{{EURASIP} J. Wireless Commun. Netw.}, vol. 2019, no.~1, pp. 1--20, May
  2019.

\bibitem{8796365}
E.~{Basar} \emph{et~al.}, ``Wireless communications through reconfigurable
  intelligent surfaces,'' \emph{IEEE Access}, vol.~7, pp. 116\,753--116\,773,
  Aug. 2019.

\bibitem{Liaskos1}
C.~{Liaskos} \emph{et~al.}, ``A new wireless communication paradigm through
  software-controlled metasurfaces,'' \emph{IEEE Commun. Mag.}, vol.~56, no.~9,
  pp. 162--169, Sept. 2018.

\bibitem{8910627}
Q.~{Wu} and R.~{Zhang}, ``Towards smart and reconfigurable environment:
  Intelligent reflecting surface aided wireless network,'' \emph{IEEE Commun.
  Mag.}, vol.~58, no.~1, pp. 106--112, Jan. 2020.

\bibitem{bariah2020prospective}
L.~Bariah \emph{et~al.}, ``A prospective look: Key enabling technologies,
  applications and open research topics in 6{G} networks,''
  \emph{arXiv:2004.06049}, 2020.

\bibitem{tariq2019speculative}
F.~Tariq \emph{et~al.}, ``A speculative study on 6{G},''
  \emph{arXiv:1902.06700}, 2019.

\bibitem{Varshney2008}
L.~R. Varshney, ``Transporting information and energy simultaneously,'' in
  \emph{Proc. IEEE Int. Symp. Inf. Theory}, Toronto, Canada, July 2008, pp.
  1612--1616.

\bibitem{Grover}
P.~Grover and A.~Sahai, ``Shannon meets tesla: Wireless information and power
  transfer,'' in \emph{Proc. IEEE Int. Symp. Int. Theory ({ISIT}'10)}, June
  2010, pp. 2363--2367.

\bibitem{Bi}
S.~Bi, C.~K. Ho, and R.~Zhang, ``Wireless powered communication: opportunities
  and challenges,'' \emph{IEEE Commun. Magazine}, vol.~53, no.~4, pp. 117--125,
  Apr. 2015.

\bibitem{Mohjazi2}
L.~Mohjazi \emph{et~al.}, ``{RF}-powered cognitive radio networks: technical
  challenges and limitations,'' \emph{IEEE Commun. Mag.}, vol.~53, no.~4, pp.
  94--100, Apr. 2015.

\bibitem{8539989}
L.~{Mohjazi} \emph{et~al.}, ``Performance analysis of {SWIPT} relaying systems
  in the presence of impulsive noise,'' \emph{IEEE Access}, vol.~6, pp.
  71\,662--71\,677, Nov. 2018.

\bibitem{8411158}
L.~{Mohjazi}, S.~{Muhaidat}, M.~{Dianati}, and M.~{Al-Qutayri}, ``Performance
  analysis of {SWIPT} relay networks with noncoherent modulation,'' \emph{IEEE
  Trans. Green Commun. Netw.}, vol.~2, no.~4, pp. 1072--1086, Dec. 2018.

\bibitem{Nasir2013}
A.~Nasir, X.~Zhou, S.~Durrani, and R.~Kennedy, ``Relaying protocols for
  wireless energy harvesting and information processing,'' \emph{IEEE Trans.
  Wireless Commun.}, vol.~12, no.~7, pp. 3622--3636, July 2013.

\bibitem{8353836}
B.~{Yang} \emph{et~al.}, ``Digital beamforming-based massive {MIMO} transceiver
  for 5{G} millimeter-wave communications,'' \emph{IEEE Trans. Microw. Theory
  Techn.}, vol.~66, no.~7, pp. 3403--3418, July 2018.

\bibitem{7374738}
E.~G. {Larsson} and H.~V. {Poor}, ``Joint beamforming and broadcasting in
  massive {MIMO},'' \emph{IEEE Trans. Wireless Commun.}, vol.~15, no.~4, pp.
  3058--3070, Apr. 2016.

\bibitem{8680660}
W.~{Hao} \emph{et~al.}, ``Beamforming design in {SWIPT}-based joint
  multicast-unicast mm{W}ave massive {MIMO} with lens-antenna array,''
  \emph{IEEE Wireless Commun. Lett.}, vol.~8, no.~4, pp. 1124--1128, Aug. 2019.

\bibitem{mohjazi2020outlook}
L.~Mohjazi \emph{et~al.}, ``An outlook on the interplay of {AI} and
  software-defined metasurfaces,'' \emph{arXiv:2004.00365}, 2020.

\bibitem{yang2016programmable}
H.~Yang \emph{et~al.}, ``A programmable metasurface with dynamic polarization,
  scattering and focusing control,'' \emph{Scientific Reports}, vol.~6, p.
  35692, Oct. 2016.

\bibitem{huang2019holographic}
C.~Huang \emph{et~al.}, ``Holographic {MIMO} surfaces for 6{G} wireless
  networks: Opportunities, challenges, and trends,'' \emph{arXiv:1911.12296},
  2019.

\bibitem{di2020smart}
M.~Di~Renzo \emph{et~al.}, ``Smart radio environments empowered by
  reconfigurable intelligent surfaces: How it works, state of research, and
  road ahead,'' \emph{arXiv:2004.09352}, 2020.

\bibitem{dunna2020scattermimo}
M.~Dunna, C.~Zhang, D.~Sievenpiper, and D.~Bharadia, ``Scatter{MIMO}: enabling
  virtual {MIMO} with smart surfaces,'' in \emph{Proc. Int. Conf. Mobile
  Computing and Networking ({M}obi{C}om'20)}, Apr. 2020, pp. 1--14.

\bibitem{246282}
V.~Arun and H.~Balakrishnan, ``{RF}ocus: Beamforming using thousands of passive
  antennas,'' in \emph{{USENIX NS}}, Feb. 2020, pp. 1047--1061.

\bibitem{ntontin2019reconfigurable}
K.~Ntontin \emph{et~al.}, ``Reconfigurable intelligent surfaces vs. relaying:
  Differences, similarities, and performance comparison,''
  \emph{arXiv:1908.08747}, 2019.

\bibitem{huang2019reconfigurable}
C.~Huang \emph{et~al.}, ``Reconfigurable intelligent surfaces for energy
  efficiency in wireless communication,'' \emph{IEEE Trans. Wireless Commun.},
  vol.~18, no.~8, pp. 4157--4170, Aug. 2019.

\bibitem{9014204}
Y.~{Yang}, S.~{Zhang}, and R.~{Zhang}, ``{IRS}-enhanced {OFDM}: Power
  allocation and passive array optimization,'' in \emph{Proc. IEEE Global
  Commun. Conf. ({GLOBECOM})}, 2019, pp. 1--6.

\bibitem{9087848}
Q.~{Nadeem} \emph{et~al.}, ``Intelligent reflecting surface assisted multi-user
  {MISO} communication: Channel estimation and beamforming design,'' \emph{IEEE
  Open J. Commun. Soc.}, vol.~1, pp. 661--680, May 2020.

\bibitem{zappone2020overhead}
A.~Zappone \emph{et~al.}, ``Overhead-aware design of reconfigurable intelligent
  surfaces in smart radio environments,'' \emph{arXiv:2003.02538}, 2020.

\bibitem{9039554}
Y.~{Yang}, B.~{Zheng}, S.~{Zhang}, and R.~{Zhang}, ``Intelligent reflecting
  surface meets {OFDM}: Protocol design and rate maximization,'' \emph{IEEE
  Trans. Commun.}, pp. 1--1, Mar. 2020.

\bibitem{8981888}
E.~{Basar}, ``Reconfigurable intelligent surface-based index modulation: A new
  beyond {MIMO} paradigm for 6{G},'' \emph{IEEE Trans. Commun.}, vol.~68,
  no.~5, pp. 3187--3196, Feb. 2020.

\bibitem{zhang2018space}
L.~Zhang \emph{et~al.}, ``Space-time-coding digital metasurfaces,''
  \emph{Nature Commun.}, vol.~9, no.~1, pp. 1--11, Oct. 2018.

\bibitem{jung2019reliability}
M.~Jung \emph{et~al.}, ``Reliability analysis of large intelligent surfaces
  ({LIS}s): Rate distribution and outage probability,'' \emph{IEEE Wireless
  Commun. Lett.}, vol.~8, no.~6, pp. 1662--1666, Dec. 2019.

\bibitem{di2019reflection}
M.~Di~Renzo and J.~Song, ``Reflection probability in wireless networks with
  metasurface-coated environmental objects: an approach based on random spatial
  processes,'' \emph{{EURASIP} J. Wireless Commun. Netw.}, vol. 2019, no.~1,
  p.~99, Apr. 2019.

\bibitem{9013789}
M.~{Jung}, W.~{Saad}, and G.~{Kong}, ``Spectral efficiency in large intelligent
  surfaces: Asymptotic analysis under pilot contamination,'' in \emph{Proc.
  IEEE Global Commun. Conf. ({GLOBECOM})}, 2019, pp. 1--6.

\bibitem{zhou2020spectral}
S.~Zhou \emph{et~al.}, ``Spectral and energy efficiency of irs-assisted {MISO}
  communication with hardware impairments,'' \emph{IEEE Wireless Commun.
  Lett.}, pp. 1--1, 2020.

\bibitem{zhang2019capacity}
S.~Zhang and R.~Zhang, ``Capacity characterization for intelligent reflecting
  surface aided {MIMO} communication,'' \emph{arXiv:1910.01573}, 2019.

\bibitem{karasik2019beyond}
R.~Karasik, O.~Simeone, M.~Di~Renzo, and S.~Shamai, ``Beyond max-{SNR}: Joint
  encoding for reconfigurable intelligent surfaces,'' \emph{arXiv:1911.09443},
  2019.

\bibitem{perovic2019channel}
N.~S. Perovi{\'c}, M.~Di~Renzo, and M.~F. Flanagan, ``Channel capacity
  optimization using reconfigurable intelligent surfaces in indoor mm{W}ave
  environments,'' \emph{arXiv:1910.14310}, 2019.

\bibitem{zhao2020performance}
W.~Zhao \emph{et~al.}, ``Performance analysis of large intelligent surface
  aided backscatter communication systems,'' \emph{IEEE Wireless Commun.
  Lett.}, 2020.

\bibitem{9027303}
V.~C. {Thirumavalavan} and T.~S. {Jayaraman}, ``{BER} analysis of
  reconfigurable intelligent surface assisted downlink power domain {NOMA}
  system,'' in \emph{Proc. Int. Conf. Commun. Syst. Netw. ({COMSNETS})}, 2020,
  pp. 519--522.

\bibitem{8847342}
Z.~{Chu}, W.~{Hao}, P.~{Xiao}, and J.~{Shi}, ``Intelligent reflecting surface
  aided multi-antenna secure transmission,'' \emph{IEEE Wireless Commun.
  Lett.}, vol.~9, no.~1, pp. 108--112, Jan. 2020.

\bibitem{zhang2020sum}
Y.~Zhang, C.~Zhong, Z.~Zhang, and W.~Lu, ``Sum rate optimization for two way
  communications with intelligent reflecting surface,'' \emph{IEEE Commun.
  Lett.}, vol.~24, no.~5, pp. 1090--1094, May 2020.

\bibitem{8811733}
Q.~{Wu} and R.~{Zhang}, ``Intelligent reflecting surface enhanced wireless
  network via joint active and passive beamforming,'' \emph{IEEE Trans.
  Wireless Commun.}, vol.~18, no.~11, pp. 5394--5409, Nov. 2019.

\bibitem{han2019intelligent}
H.~Han \emph{et~al.}, ``Intelligent reconfigurable surface aided power control
  for physical-layer broadcasting,'' \emph{arXiv:1912.03468}, 2019.

\bibitem{zhou2019robust}
G.~Zhou \emph{et~al.}, ``Robust beamforming design for intelligent reflecting
  surface aided {MISO} communication systems,'' \emph{arXiv:1911.06237}, 2019.

\bibitem{8000613}
H.~{Lang} and C.~D. {Sarris}, ``Optimization of wireless power transfer systems
  enhanced by passive elements and metasurfaces,'' \emph{IEEE Trans. Antennas
  Propag.}, vol.~65, no.~10, pp. 5462--5474, Oct. 2017.

\bibitem{8941080}
Q.~{Wu} and R.~{Zhang}, ``Weighted sum power maximization for intelligent
  reflecting surface aided {SWIPT},'' \emph{IEEE Wireless Commun. Lett.},
  vol.~9, no.~5, pp. 586--590, May 2020.

\bibitem{pan2020intelligent}
C.~Pan \emph{et~al.}, ``Intelligent reflecting surface aided {MIMO}
  broadcasting for simultaneous wireless information and power transfer,''
  \emph{IEEE J. Sel. Areas Commun.}, 2020.

\bibitem{wu2019joint}
Q.~Wu and R.~Zhang, ``Joint active and passive beamforming optimization for
  intelligent reflecting surface assisted {SWIPT} under {Q}o{S} constraints,''
  \emph{arXiv:1910.06220}, 2019.

\bibitem{6952122}
D.~{Altinel} and G.~K. {Kurt}, ``Statistical models for battery recharging time
  in rf energy harvesting systems,'' in \emph{Proc. IEEE Wireless Commun. Netw.
  Conf. ({WCNC})}, 2014, pp. 636--641.

\bibitem{7374745}
D.~{Altinel} and G.~{Karabulut Kurt}, ``Energy harvesting from multiple {RF}
  sources in wireless fading channels,'' \emph{IEEE Trans. Veh. Technol.},
  vol.~65, no.~11, pp. 8854--8864, Nov. 2016.

\bibitem{8269106}
E.~{Salahat} and N.~{Yang}, ``Statistical models for battery recharge time from
  {RF} energy scavengers in generalized wireless fading channels,'' in
  \emph{Proc. IEEE Globecom Workshops ({GC W}kshps)}, 2017, pp. 1--6.

\bibitem{qian2020beamforming}
X.~Qian \emph{et~al.}, ``Beamforming through reconfigurable intelligent
  surfaces in single-user {MIMO} systems: {SNR} distribution and scaling laws
  in the presence of channel fading and phase noise,'' \emph{arXiv:2005.07472},
  2020.

\bibitem{tang2019wireless}
W.~Tang \emph{et~al.}, ``Wireless communications with reconfigurable
  intelligent surface: Path loss modeling and experimental measurement,''
  \emph{arXiv:1911.05326}, 2019.

\bibitem{di2020analytical}
M.~Di~Renzo \emph{et~al.}, ``Analytical modeling of the path-loss for
  reconfigurable intelligent surfaces--anomalous mirror or scatterer?''
  \emph{arXiv:2001.10862}, 2020.

\bibitem{Prudnikov}
A.~P. Prudnikov, Y.~A. Brychkov, and O.~I. Marichev, \emph{Integrals and
  Series}.\hskip 1em plus 0.5em minus 0.4em\relax Gordon and Breach Science
  Publishers, 1986, vol.~3.

\bibitem{Abramowitz}
M.~Abramowitz, \emph{Handbook of Mathematical Functions, With Formulas, Graphs,
  and Mathematical Tables,}.\hskip 1em plus 0.5em minus 0.4em\relax USA: Dover
  Publications, Inc., 1974.

\bibitem{8245828}
F.~{El Bouanani} and D.~B. {da Costa}, ``Accurate closed-form approximations
  for the sum of correlated {W}eibull random variables,'' \emph{IEEE Wireless
  Commun. Lett.}, vol.~7, no.~4, pp. 498--501, Aug. 2018.

\bibitem{LinaB}
L.~Bariah, S.~Muhaidat, P.~C. Sofotasios, and F.~El-bouanani, ``Large
  intelligent surfaces-based non-orthogonal multiple access: Performance
  analysis,'' \emph{in submission}, 2020.

\bibitem{wolfram}
\BIBentryALTinterwordspacing
 [Online]. Available: \url{http://functions.wolfram.com/}
\BIBentrySTDinterwordspacing

\bibitem{papoulis}
A.~Papoulis, \emph{Probability, Random Variables, and Stochastic
  Processes}.\hskip 1em plus 0.5em minus 0.4em\relax New York: McGraw-Hill, 3rd
  edition, 1991.

\bibitem{Simon}
M.~K. Simon and M.-S. Alouini, \emph{Digital Communications over Fading
  Channels. A Unified Approach to Performance Analysis.}\hskip 1em plus 0.5em
  minus 0.4em\relax New York, NY: John Wiley and Sons, Inc., 2000.

\bibitem{Proakis}
J.~G. Proakis, \emph{Digital Communications}.\hskip 1em plus 0.5em minus
  0.4em\relax New York: McGraw-Hill, 4th edition, 2000.

\end{thebibliography}
\end{document}